\documentclass[journal]{IEEEtran}
\usepackage{cite}
\ifCLASSINFOpdf
   \usepackage[pdftex]{graphicx}
\else
\fi
\usepackage{amsmath}
\usepackage{amsfonts}
\usepackage{setspace}

\usepackage{soul} 
\usepackage{booktabs} 

\usepackage[colorlinks=true,linkcolor=blue,citecolor=blue,urlcolor=black ]{hyperref}

\usepackage{graphicx}

\usepackage{subcaption}

\begin{document}
%
% paper title
% Titles are generally capitalized except for words such as a, an, and, as,
% at, but, by, for, in, nor, of, on, or, the, to and up, which are usually
% not capitalized unless they are the first or last word of the title.
% Linebreaks \\ can be used within to get better formatting as desired.
% Do not put math or special symbols in the title.
\title{Multi-image Super Resolution of Remotely Sensed Images using Residual Feature Attention Deep Neural Networks}
%
%
% author names and IEEE memberships
% note positions of commas and nonbreaking spaces ( ~ ) LaTeX will not break
% a structure at a ~ so this keeps an author's name from being broken across
% two lines.
% use \thanks{} to gain access to the first footnote area
% a separate \thanks must be used for each paragraph as LaTeX2e's \thanks
% was not built to handle multiple paragraphs
%

\author{Francesco~Salvetti,
        Vittorio~Mazzia, Aleem~Khaliq,
        and~Marcello~Chiaberge% <-this % stops a space
\thanks{The authors are with Politecnico di Torino -- Department of Electronics and Telecommunications, PIC4SeR, Politecnico di Torino Interdepartmental Centre for Service Robotics and SmartData@PoliTo, Big Data and Data Science Laboratory, Italy. Email: \{name.surname\}@polito.it.}}% <-this % stops a space

% note the % following the last \IEEEmembership and also \thanks - 
% these prevent an unwanted space from occurring between the last author name
% and the end of the author line. i.e., if you had this:
% 
% \author{....lastname \thanks{...} \thanks{...} }
%                     ^------------^------------^----Do not want these spaces!
%
% a space would be appended to the last name and could cause every name on that
% line to be shifted left slightly. This is one of those "LaTeX things". For
% instance, "\textbf{A} \textbf{B}" will typeset as "A B" not "AB". To get
% "AB" then you have to do: "\textbf{A}\textbf{B}"
% \thanks is no different in this regard, so shield the last } of each \thanks
% that ends a line with a % and do not let a space in before the next \thanks.
% Spaces after \IEEEmembership other than the last one are OK (and needed) as
% you are supposed to have spaces between the names. For what it is worth,
% this is a minor point as most people would not even notice if the said evil
% space somehow managed to creep in.

% The paper headers
\markboth{}%
{Francesco Salvetti \MakeLowercase{\textit{et al.}}: Multi-image Super Resolution of Remotely Sensed Images using Residual Feature Attention Deep Neural Networks}
% The only time the second header will appear is for the odd numbered pages
% after the title page when using the twoside option.
% 
% *** Note that you probably will NOT want to include the author's ***
% *** name in the headers of peer review papers.                   ***
% You can use \ifCLASSOPTIONpeerreview for conditional compilation here if
% you desire.

% If you want to put a publisher's ID mark on the page you can do it like
% this:
%\IEEEpubid{0000--0000/00\$00.00~\copyright~2015 IEEE}
% Remember, if you use this you must call \IEEEpubidadjcol in the second
% column for its text to clear the IEEEpubid mark.

% use for special paper notices
%\IEEEspecialpapernotice{(Invited Paper)}

% make the title area
\maketitle

% As a general rule, do not put math, special symbols or citations
% in the abstract or keywords.
\begin{abstract}
Convolutional Neural Networks (CNNs) have been consistently proved state-of-the-art results in image Super-Resolution (SR), representing an exceptional opportunity for the remote sensing field to extract further information and knowledge from captured data. However, most of the works published in the literature have been focusing on the Single-Image Super-Resolution problem so far. At present, satellite based remote sensing platforms offer huge data availability with high temporal resolution and low spatial resolution. In this context, the presented research proposes a novel residual attention model (RAMS) that efficiently tackles the multi-image super-resolution task, simultaneously exploiting spatial and temporal correlations to combine multiple images. We introduce the mechanism of visual feature attention with 3D convolutions in order to obtain an aware data fusion and information extraction of the multiple low-resolution images, transcending limitations of the local region of convolutional operations. Moreover, having multiple inputs with the same scene, our representation learning network makes extensive use of nestled residual connections to let flow redundant low-frequency signals and focus the computation on more important high-frequency components. Extensive experimentation and evaluations against other available solutions, either for single or multi-image super-resolution, have demonstrated that the proposed deep learning-based solution can be considered state-of-the-art for Multi-Image Super-Resolution for remote sensing applications.
\end{abstract}

% Note that keywords are not normally used for peerreview papers.
\begin{IEEEkeywords}
Deep Learning, Multi-Image Super-resolution, Attention Networks, 3D Convolutional Neural Networks
\end{IEEEkeywords}

% For peer review papers, you can put extra information on the cover
% page as needed:
% \ifCLASSOPTIONpeerreview
% \begin{center} \bfseries EDICS Category: 3-BBND \end{center}
% \fi
%
% For peerreview papers, this IEEEtran command inserts a page break and
% creates the second title. It will be ignored for other modes.

\section{Introduction}
Super-resolution (SR) algorithms serve the purpose of reconstructing high-resolution (HR) images from either single or multiple low-resolution (LR) images. Due to constraints such as sensor limitations and exceedingly high acquisition costs, it is often challenging to obtain HR images. In this regard, SR algorithms provide viable opportunity to enhance and reconstruct HR images from LR images recorded by the sensors. Over more than three decades, progress has steadily been observed in the development of Super-resolution, as both multi-frame and single-frame SR now have substantial applications that can use the image generation purposefully.

SR is very significant to Remote Sensing because it provides opportunity to enhance LR images despite the inherent problems often faced in remote-sensing scenarios. The hardware and material costs for smaller missions around data accumulation are very high. Additionally, onboard instruments on satellites continue to generate ever-increasing data as spatial and spectral resolutions also increase, and this has progressively become challenging for compression algorithms \cite{valsesia2014novel}, as they try to meet the bandwidth restrictions \cite{benecki2018evaluating,valsesia2016universal}. 
Remote sensing is fundamental in obtaining images covering most of the globe, permitting many vital projects such as disaster monitoring, military surveillance, urban maps, and vegetation growth monitoring. It is thus imperative that enhancements and progress be made in post-processing techniques to overcome obstacles of increasing spatial resolution.

There are two main methods used in Super-resolution: Single-image SR (SISR) and Multi-image SR (MISR). SISR employs a single image to reconstruct a HR version of it. However, a single image is quite limited in the amount of information that it provides, mainly post the LR image formation process. Contrastingly, MISR involves multiple LR images of the same scene acquired from the same or different sensors to construct an HR image. The significant advantage MISR holds over SISR is in how it can draw out otherwise unavailable information from the different image observations of the same scene. It consequently constructs high spatial resolution image. 
However, to achieve the additional benefits of MISR, a multitude of problems need to be solved. Conventionally, multiple images are obtained by either a satellite during its multiple orbits or by different satellites at different times or different sensors acquiring images at the same time. With so many variables involved, many complications need to be considered, such as cloud coverage, time variance in scene content, and invariance to absolute brightness variability.

There has been significant progress in Single-image SR as deep learning methods and deep neural networks have been brought into use, allowing a better efficient generation of non-linear maps to deal with complex degradation models. However, there has not been any similar progress in MISR.\\

In this paper, building over the latest breakthroughs in SISR \cite{dong2015image, lim2017enhanced, yu2018wide, zhang2018image, dai2019second}, we propose a deep learning MISR solution for remote-sensing applications that exploits both spatial and temporal correlations to combine multiple low-resolution acquisitions smartly. Indeed, our model provides a real end-to-end efficient solution to recover high-resolution images, overcoming limitations of previous similar methodologies, and providing enhanced reconstruction results. So, the presented research constitutes an exceptional opportunity, easily replicable, to access better quality and more useful information for the remote-sensing community. In particular, the main contribution of our work lies in:
\begin{enumerate}
\item The use of 3D convolutions to efficiently extract, directly from the stack of multiple low-resolution images, high-level representations, simultaneously exploiting spatial and temporal correlations.
\item The introduction of a novel feature attention mechanism for 3D convolutions that lets the network focus on most promising high-frequency information largely overcoming main locality limitations of convolutional operations. Moreover, the concurrent use of multiple nested residuals, inside the network, let low-frequency components flow directly to the output of the model.
\item The conceptualization and development of an efficient, highly replicable, deep learning neural network for MISR that makes use of 2D and 3D convolutions exclusively in the low-resolution space. It has been extensively evaluated on a major multi-frame open-source remote-sensing dataset proving state-of-the-art results with a considerable margin. So, it constitutes an exceptional tool and opportunity for the remote-sensing research community.
\end{enumerate}

The remainder of this paper is structured as follows. Section \ref{relatedwork} covers the related work on SR and its developments in techniques for both SISR and MISR. Section \ref{methodology} explains the overall methodology, network architecture and its subsequent blocks, and training process. Section \ref{experiments} discusses the experimentation, the Proba-V dataset, data pre-processing, and results. Section \ref{conclusion} draws some conclusions and future directions.

\section{Related Work} \label{relatedwork}
Related literature is organized as follows. Firstly, a wide range of studies related to SISR are discussed which involve state-of-the-art methods and recent developments in SISR techniques, which is the basis of every SR method. Secondly, studies performed for SR in remote sensing domain are discussed. Lastly, MISR related studies, which are rarely addressed, are discussed including latest developments.     
\subsection{Single-image Super-resolution}
Ever since the late eighties and the early nineties, there has been an eager interest in SR, comprehensively reviewed by Borman and Stevenson \cite{borman1998super}. Following forth in the works of Tsai and Huang \cite{tsai1984multiframe} and afterward, Kim et al. \cite{kim1990recursive}, the new approaches considered processing images in the frequency domain to recover lost information of higher-frequency. These first works had certain drawbacks, like the level of difficulty observed in successfully incorporating the prior available spatial information. Several studies performed by Irani and Peleg \cite{irani1990super,irani1991improving,irani1993motion} focused over the spatial domain, proposing methods for SR reconstruction.\\
Learning-based methods build upon the relation between LR-HR images, and there have been many recent advancements in this approach, mostly due to deep convolutional neural networks (CNNs) \cite{dong2016accelerating,kim2016accurate,dong2015image}. The leading force for this was Dong et al. \cite{dong2014learning}, who achieved superior results by proposing a Super-resolution CNN (SRCNN) framework. Kim et al. introduced residual learning and suggested very deep SR (VDSR) \cite{kim2016accurate} and deeply recursive CN (DRCN) \cite{kim2016deeply} with 20 layers. Later, Tai et al. pioneered deep recursive residual network (DRRN) \cite{tai2017image} and memory blocks in MemNet \cite{tai2017memnet}. So going forth, particular emphasis has been placed on proper upscaling of spatial resolutions at network tail-ends, as well as extracting information of the original scale LR inputs. To that end, some enhancements have been proposed for accelerating the testing and training needed for SRCNN, a faster network structure FSRCNN \cite{dong2016accelerating}. Ledig et al. \cite{ledig2017photo} proposed SRGAN, a generative adversarial network (GAN) for photo-realistic SR with perceptual losses \cite{johnson2016perceptual}, and  K. He et al. introduced ResNet \cite{he2016deep} for image SR and to make a deeper network SRResNet. EnhanceNet \cite{sajjadi2017enhancenet} also used a GAN based model to merge perceptual loss with automated texture synthesis.  Though, the predicted results can produce some artifacts and may not be a faithful reconstruction. 

In recent past years, enhancements in deep networks have been proposed and showed promising results for SISR, for example, in \cite{lim2017enhanced}, an Enhanced Deep Super-resolution (EDSR) network was developed to improve the performance by removing unnecessary modules and expanding the model size with the stable training process in conventional residual networks. Yu et. al \cite{yu2018wide} demonstrated better results in terms of accurate SR by generating models with a wide range of features before ReLU activation and training with normalized weights. Zhang et. al \cite{zhang2018image} proposed residual channel attention networks (RCAN) that exploits very deep network structure based on residual in residual (RIR) which bypass excessive low-frequency information through multiple skip connections. 
\subsection{SR for Remotely Sensed Imagery}
With the increasing availability of recent satellite-based multispectral sensors and transmission bandwidth restrictions \cite{lanaras2018super}, it is possible to obtain images at different spatial resolutions with multiple spectral bands. Keen attention is being paid to developing better methods of super-resolving the lower-resolution bands but simultaneously keeping the images at a high spatial resolution. An example can be seen in \cite{liebel2016single}, where - through the utilization of only lower resolution bands – SR of multispectral remote sensing images is applied with convolutional layers. \cite{lei2017super} shows the integration of residual connections into a single image SR based architecture to achieve better SR performance. The performance of image enhancement methods in computer vision can also be increased prominently through generative adversarial networks (GANs) \cite{ledig2017photo,wang2018esrgan}. Moreover,  GANs have also been exploited to super-resolve remote sensing images. For example, Ma et. al. \cite{ma2019super} developed a dense residual generative adversarial network (DRGAN)-based SISR method to super resolve remote sensing images.  By designing a dense residual network as the generative network in GAN, their method makes full use of the hierarchical features from low-resolution (LR) images. \\
Dong et. al. \cite{dong2019transferred} proposed a novel multi-perception attention network (MPSR) for Super-resolution of low resolution remotely sensed images, which achieved better results by incorporating the proposed enhanced residual block (ERB) and residual channel attention group (RCAG). Their methodology is capable of dealing with low-resolution remote sensing images via multi-perception learning and multi-level information adaptive weighted fusion. They claimed that, a pre-train and transfer learning strategy can improve the SR performance and stabilize the training procedure. Gargiulo et. al. \cite{gargiulo2019fast} proposed a CNNs based approach to provide a 10 m super-resolved image of the original 20 m bands of remotely sensed Sentinel-2 images. In their experimental results,  they claimed  that the proposed solution can achieve better performance with respect to most of the state-of-the-art methods, including other deep learning based ones with a considerable saving of computational burden. Recently methods to enhance spatial resolution of remotely sensed images used Parallel Residual Network \cite{wu2020sentinel}, Bidirectional Convolutional LSTMs \cite{chang2019bidirectional}, Deep Residual Squeez and Excitation Network \cite{gu2019deep}.

\subsection{Multi-image Super-resolution}
Multi-image SR (MISR) involves the extraction of information from many LR observations of the same scene to reconstruct HR images \cite{yue2016image}. The earliest work for MISR was proposed by Tsai and Huang \cite{tsai1984multiframe} using a frequency-domain technique, by combining multiple images with sub-pixel displacements to improve the spatial resolution of images. Due to the some weaknesses of the first proposed method related to incorporate prior information of HR images, several spatial domain MISR techniques were considered \cite{elad2001fast}. These include projection onto convex sets (POCS) \cite{stark1989high}, non-uniform interpolation \cite{lertrattanapanich2002high}, regularized methods \cite{takeda2007kernel,shen2009super}, and sparse coding \cite{kato2017double}. With the availability of more data from the multiple observations of the scene, it is possible to obtain a more accurate reconstruction than through single-image methods. MISR techniques involve different ways of degrading the original image by following an image model, and these involve blurring, warping, noise contamination, and decimation. Then the degradation is reversed by solving an ill-posed optimization problem. To this end, Bayesian reconstruction in the gradient projection algorithm was used alongside subpixel displacement estimation \cite{schultz1996extraction}. An enhanced Fast and Robust SR (FRSR) \cite{farsiu2004fast} employs estimation of maximum likelihood analysis and simplified regulation. Another proposal in SR was for the Adaptive detail enhancement (SR-ADE) \cite{arbelaez2010contour}, which reconstructs satellite images with the use of a bilateral filter for decomposing input images while also amplifying high-frequency detail information. 

Another approach Iterative Back Projection (IBP), introduced by Irani and Peleg \cite{irani1991improving}, used a back-projection of the difference between the actual LR images obtained and the simulated LR images to the SR image. The forward imaging process is inverted and iteratively attempted in updates. As with MISR, there are apparent drawbacks when prior images are difficult to be included, or it is difficult to model an image's degradation process.

In the past few years, many deep learning-based approaches have been exploited to address the MISR problems in the context of enhancing video sequences \cite{kappeler2016video,caballero2017real,jo2018deep}. However, MISR is rarely exploited for remotely sensed satellite imagery. Kawulok et al \cite{kawulok2019deep} demonstrated the potential benefits of information fusion offered by multiple satellite images reconstruction and learning-based SISR approaches. In their work, EvoNet framework \cite{kawulok2018evolving} based on several deep CNNs was adopted to exploit SISR in the preprocessing phase of the input data for MISR. 

Recently, a challenge was set by the European Space Agency (ESA) to super-resolved multi-temporal PROBA-V satellite imagery\footnote{https://kelvins.esa.int/proba-v-super-resolution.}. In this context, a new CNN-based architecture \textit{DeepSUM} was proposed by Molini et. al \cite{molini2019deepsum} to super resolve multi-temporal PROBA-V imagery. An end-to-end learning approach was established by exploiting both spatial and temporal correlations.
Most recently, Deudon et. al presented \textit{HighRes-net} based on deep learning to deal with the MISR of remotely sensed PROBA-V satellite imagery \cite{deudon2020highres}. They proposed an end-to-end mechanism that learns the sub-tasks involved in MISR, that are co-registration, fusion, upsampling, and registration-at-the-loss.
%----------- 

%----------- Vitto's Work
\section{Methodology} \label{methodology}
MISR aims at recovering an HR image $I^{\text{HR}}$ from a set of $T$ LR images $I^{\text{LR}}_{[1,T]}$ of the same scene acquired in a certain temporal window. In contrast to SISR, MISR can simultaneously benefit from spatial and temporal correlations, being able to achieve far better reconstruction results theoretically. Either way, SR is an inherently ill-posed problem since a multiplicity of solutions exist for any given set of low-resolution images. So, it is an underdetermined inverse problem, of which solution is not unique. Our proposed methodology, based on a representation learning model, aims at generating a super-resolved image $I^{\text{SR}}$ applying a function $H_\text{RAMS}$ to the set of $I^{\text{LR}}_{[1,T]}$ images:
\begin{equation}
I^{\text{SR}}=H_\text{RAMS}(I^{\text{LR}}_{[1,T]}, \Theta)
\end{equation}
where $\Theta$ are model parameters learned with an iterative optimization process.

In other words, we propose a fully convolutional Residual Attention Multi-image Super-resolution network (RAMS) that can efficiently extract high-level features concurrently from $T$ LR images and fuse them exploiting a built-in visual attention mechanism. Attention directs the focus of the model only on most promising extracted features, reducing the importance of less relevant ones and mostly transcending limitations of the local region of convolutional operations. Moreover, extensive use of nested residual connections lets all the redundant low-frequency information, present in the set $I^{\text{LR}}_{[1,T]}$ of LR images, flow through the network, leaving the model focusing its computation only on high-frequency components. Indeed, high-frequency features are more informative for HR reconstruction, while LR images contain abundant low-frequency information that can directly be forwarded to the final output \cite{zhang2018image}. Finally, as the majority of the model for single-image super-resolution \cite{lim2017enhanced, yu2018wide, dong2016accelerating, dai2019second}, all computations in our network are efficiently performed in the LR feature space requiring only an upsample operation at the final stage of the model. 

In the following paragraphs, we present the overall architecture of the network with a detailed overview of the main blocks. Finally, we conclude the methodology section with precise details of the optimization process for training the network. 

\subsection{Network architecture}
\begin{figure*}[h]
\centering
\includegraphics[scale=0.32]{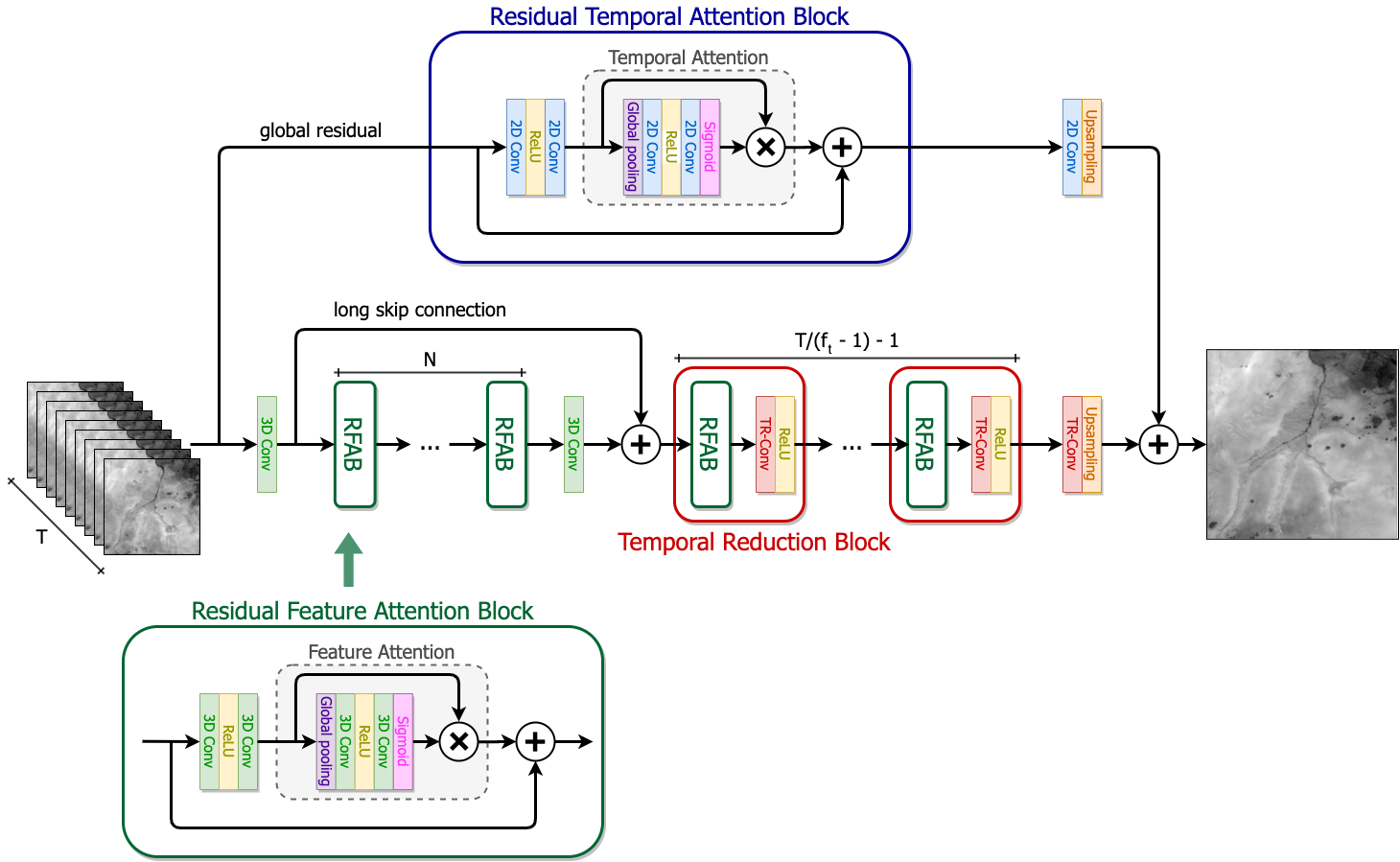}
\caption{Overview of the Residual Attention Multi-image Super-resolution Network (RAMS), assuming to work with single-channel LR images ($C=1$) to simplify the discussion. A tensor of $T$ single-channel LR images constitutes the input of the proposed model. The main branch extracts features, with 3D convolutions, in a hierarchical fashion, while a feature attention mechanism allows the network to select and focus on most promising inner representations. Concurrently, a global residual path exploits a similar attention operation in order to make an aware fusion of the $T$ distinct LR images. All computations are efficiently performed in the LR feature space and only at the last stage of the model an upsampling operation is performed in both branches.}
\label{fig:RAMS_architecture}
\end{figure*}
An overview of the RAMS network, with its main three blocks and two branches, is depicted in Fig. \ref{fig:RAMS_architecture}. As a high-level description, the model takes as input a single set of $T$ low-resolution images $I^{\text{LR}}_{[1,T]}$ that can be represented as a tensor $\mathrm{\textbf{X}}^{(i)}$ with shape $H\times W\times T \times C$ where $H$, $W$ and $C$ are the height, width, and channels of the single low-resolution images, respectively. The upper global residual path proposes a simple SR solution, making an aware fusion of the $T$ input images. On the other hand, the central branch exploits 3D convolutions residual-based blocks in order to extract spatial and temporal correlations from the same set of $T$ LR images and provide a refinement to the residual simple SR image.

More in detail, in the first part of the main path of the model, we use a simple 3D convolutional layer, with each filter of size $f_{h}\times f_{w}\times f_{t}$, to extract $F$ shallow features from the input set $I^{\text{LR}}_{[1,T]}$ of LR images. Then, we apply a cascade of N residual feature attention blocks that increasingly extract higher-level features, exploiting local and non-local, temporal, and spatial correlations. Moreover, we make use of a long skip connection for the shallow features and several short skip connections inside each feature attention block to let flow all redundant low-frequency signals and let the network focus on more valuable high-frequency components. The three dimensions $H$, $W$ and $T$ are always preserved through reflecting padding. The first part of the main branch can be modeled as a single function $H_{I}$ that maps each tensor $\mathrm{\textbf{X}}^{(i)}$ to a new higher dimensional one $\mathrm{\textbf{X}}^{(i)}_{I}$ with shape $H\times W\times T\times F$:
\begin{equation}
\mathrm{\textbf{X}}^{(i)}_{I} = H_{I}(\mathrm{\textbf{X}}^{(i)})
\end{equation}

In the second part of the main branch, we further process the output tensor $\mathrm{\textbf{X}}^{(i)}_{I}$ with $\lfloor T/(f_{t}-1) \rfloor -1$ temporal reduction blocks. In each block, we intersperse a residual feature attention block with 3D convolutional layer without padding on the temporal $T$ dimension (TR-Conv). So, $H$, $W$ and $F$ remain invariant and only the temporal dimension is reduced. The output of this second block is a new tensor $\mathrm{\textbf{X}}^{(i)}_{II}$ with shape $H\times W\times f_{t}\times F$, where the temporal dimension $T$ is reduced to $f_{t}$:
\begin{equation}
\mathrm{\textbf{X}}^{(i)}_{II} = H_{II}(\mathrm{\textbf{X}}^{(i)}_{I})
\end{equation}

Finally, the output tensor $\mathrm{\textbf{X}}^{(i)}_{II}$ is processed by a further TR-Conv layer that reduces $T$ to one and an upscale function $H_{\text{UP}|_\text{3D}}$ that generates a tensor $\mathrm{\textbf{X}}^{(i)}_{\text{UP}|_\text{3D}}$ of shape $sH\times sW\times C$ where $s$ is the scaling factor.
\begin{figure*}
\centering
\includegraphics[scale=0.57]{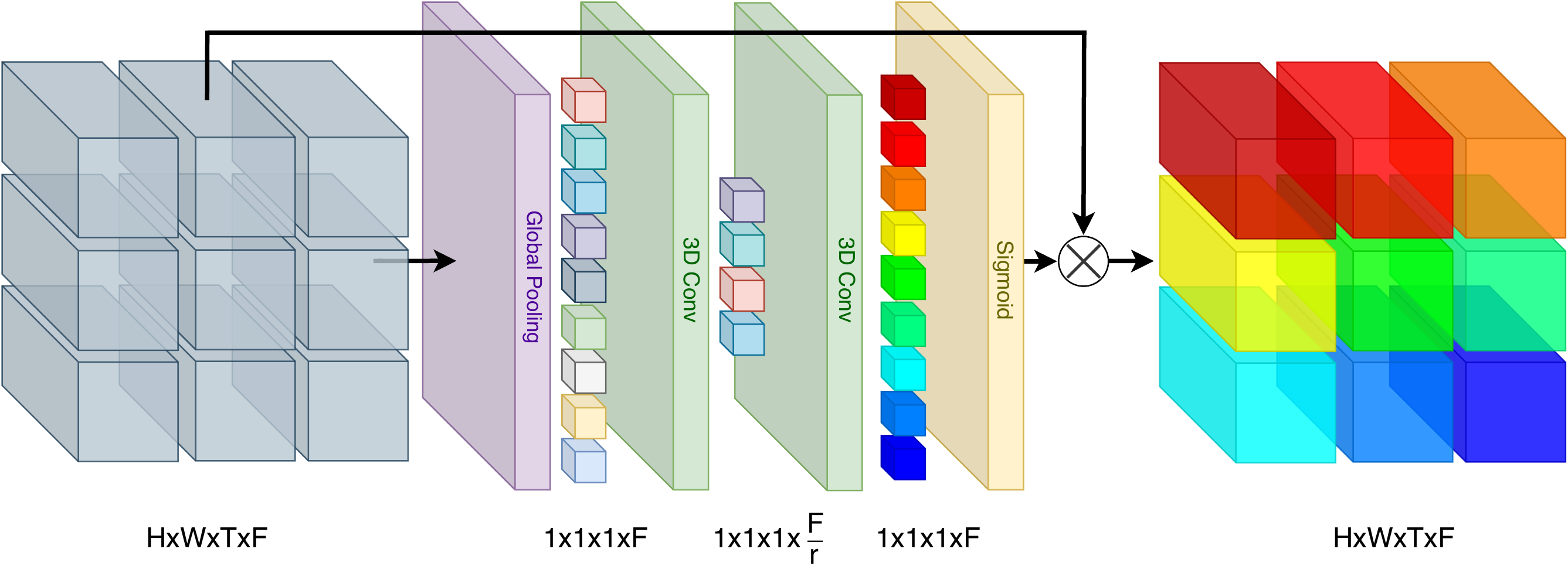}
\caption{Reference architecture of a feature attention block. A series of convolutional operations and non-linear activations are applied to the input tensor with shape $H \times W \times T \times F$ in order to generate different attention statistics for each feature $F$ that concurrently take advantage of local and non-local correlations. Consequently, each tensor's feature is properly re-scaled, enabling the network to focus on most promising components and letting residual connections heed of all redundant low-frequency signals.}
\label{fig:feature_attention}
\end{figure*}
The overall output $\mathrm{\textbf{X}}^{(i)}_{\text{UP}|_\text{3D}}$ of the main branch sums with the trivial solution provided by the global residual. Indeed, the global path simply weights the $T$ LR images of the input tensor $\mathrm{\textbf{X}}^{(i)}$ with a residual temporal attention block with filters of size $f_{h}\times f_{w}$. Then it produces an output tensor $\mathrm{\textbf{X}}^{(i)}_{\text{UP}|_\text{2D}}$ of shape $sH\times sW\times C$ that is added to the one of the main branch.  So, the final SR prediction of the network $\mathrm{\hat{\textbf{Y}}}^{(i)}=I^{\text{SR}}$ is the sum of the two contribution:
\begin{equation}
\mathrm{\hat{\textbf{Y}}}^{(i)}=
H_\text{RAMS}(\mathrm{\textbf{X}}^{(i)})= (\mathrm{\textbf{X}}^{(i)}_{\text{UP}|_\text{3D}} + \mathrm{\textbf{X}}^{(i)}_{\text{UP}|_\text{2D}})
\end{equation}

The upscaling procedure is identical for both branches; 
after several trials with different methodologies, such as transposed convolutions \cite{deudon2020highres}, bi-linear resizing and nearest-neighbor upsampling \cite{van2017learned}, we adopted a sub-pixel convolution layer as explained in detail in \cite{shi2016real}. So, for either branch, the last 2D or 3D convolution generates $s^{2}\cdot C$ features in order to produce the final tensors of shape $sH\times sW\times C$ for the residual sum.

In conclusion, the overall model takes as input a tensor $\mathrm{\textbf{X}}^{(i)}$ with shape $H\times W\times T \times C$, works always efficiently in the LR space and generates only at the final stage an output tensor $\mathrm{\hat{\textbf{Y}}}^{(i)}$ with shape $sH\times sW\times C$. 

In the following sub-paragraphs, the three major functional blocks, residual feature attention, residual temporal attention, and temporal reduction blocks are further explained and analyzed.

\subsection{Residual attention blocks}
Residual attention blocks are at the core of the RAMS model; their specific architecture allows it to focus on relevant high-frequency components and let redundant, low-frequency information flow through the residual connections of the network.
Inter-dependencies among features, in the case of feature attention blocks, or temporal steps, in the case of temporal attention blocks, are taken into account computing for each of them, relevant statistics that take into account local and non-local, temporal and spatial correlations. Indeed, either 3D or 2D convolution filters operate with local receptive fields loosing the possibility to exploit contextual information outside of their limited region of view.
\subsubsection{Residual feature attention}
Except for the global residual path, all residual attention blocks are residual feature attention blocks, as shown in Fig. \ref{fig:RAMS_architecture}. Indeed, each block of features is weighted up in order to trace most promising high-frequency components, and a residual connection lets low-frequency information flow through the network.

More formally, the output of a residual feature attention block with a generic tensor, $\mathrm{\textbf{X}}^{(i)}_{n}$, is equal to:
\begin{equation}
F_{RFA}(\mathrm{\textbf{X}}^{(i)}_{n})=\mathrm{\textbf{X}}^{(i)}_{n} + H_{FA}(\mathrm{\textbf{X}}^{(i)}_{*})\cdot\mathrm{\textbf{X}}^{(i)}_{*}
\label{residual_feature_attention}
\end{equation}
where $H_{FA}$ is the feature attention function and $\mathrm{\textbf{X}}^{(i)}_{*}$ is the output of two stacked 3D convolutional layers.
\begin{equation}
\mathrm{\textbf{X}}^{(i)}_{*}= W_{2}\ast max(0, W_{1}\ast\mathrm{\textbf{X}}^{(i)}_{n} + B_{1}) + B_{2}
\end{equation}
where $W_{1}$,$W_{2}$ and $B_{1}$, $B_{2}$ represent the filters with size $f_{h}\times f_{w}\times f_{t}$ and biases respectively and, '$\ast$' denotes the 3D convolution operation.
The number of filters is always equal to $F$ as the ones extracted by the first 3D convolutional layer. 

So, all low-frequency components in $\mathrm{\textbf{X}}^{(i)}_{n}$ can flow through the residual connection and $H_{FA}$ can focus the attention of the network to more valuable high-frequency signals. 
To this end, the feature attention block takes the feature-wise global spatial and temporal information into a feature descriptor by using a global average pooling. So, from the tensor $\mathrm{\textbf{X}}^{(i)}_{*}$ with shape $H\times W\times T\times F$ we extract $z_{F}\in \mathbb{R}^{F}$ feature statistics using the following equation:
\begin{equation}
z_{F}=\frac{1}{H\times W\times T}\sum_{i=1}^{H} \sum_{j=1}^{W}\sum_{k=1}^{T}\mathrm{\textbf{X}}^{(i)}_{*}(i,j,k)
\end{equation}
Statistics of the feature $z_{F}$ can be viewed as a collection of descriptors, whose values contribute to express the whole stack of temporal images \cite{hu2018squeeze}.

In Fig.\ref{fig:feature_attention}, it is possible to observe the global pooling operation which output is a tensor $\mathrm{\textbf{Z}}^{(i)}_{F}$ with shape $1\times1\times1\times F$ and last dimension equal to $z_{F}$. In addition, the output tensor $\mathrm{\textbf{Z}}^{(i)}_{F}$ is further processed by a stack of two 3D convolutional layers with a ReLU \cite{nair2010rectified} and sigmoid activation function, respectively. Indeed, as discussed in \cite{hu2018squeeze}, the stack of two convolutional layers with the filter of size $1\times 1\times 1$ concur to create a non-linear mapping function which is able to deeply capture feature-wise dependencies from the aggregated information extracted by the global pooling operation. The first 3D convolutional layer reduces the feature size by a factor of $r$, and then the second layer restores the original dimension and constraints its values from zero to one with a sigmoid function in a non-mutually exclusive relationship.

Finally, the original tensor $\mathrm{\textbf{X}}^{(i)}_{*}$ is weighted up by the processed attention statistics as shown in Eq. \ref{residual_feature_attention}.
\subsubsection{Residual temporal attention}
\begin{figure}[b]
\includegraphics[scale=0.45]{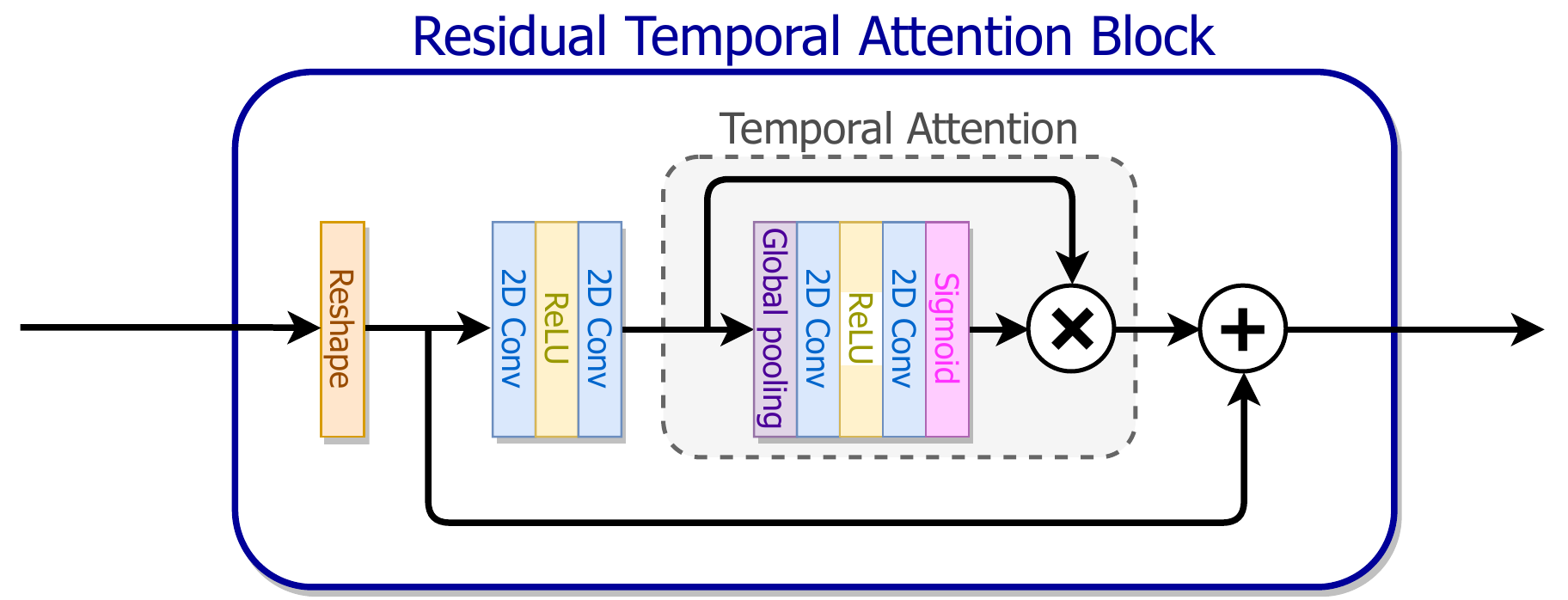}
\caption{Reference architecture of a residual temporal attention block. If the number of channels $C\neq 1$ the input tensor $\mathrm{\textbf{X}}^{(i)}$ is reshaped in $H\times W\times (T\cdot C)$. Consequently, all temporal channels are weighted with some statistics computed by the layers of the temporal attention block.}
\label{fig:temporal_attention}
\end{figure}
The primary purpose of the global residual path is to generate a starting trivial solution for the upsampling problem. More accurate is this starting prediction, and more simplified is the role of the main branch of the network, leading to a lower reconstruction error. However, the input of the model $\mathrm{\textbf{X}}^{(i)}$ has $T$ different LR images that have to be merged. Intuitively, for each input sample $I^{\text{LR}}_{[1,T]}$, there are some LR images more similar to each other. So, giving them more relevance when merging the $T$ LR images would most probably lead to higher quality predictions. In this context, the aim of the residual temporal attention block is to make an aware weighing of the different input temporal images, letting the network to make an upsample solution with primarily the most similar temporal steps.
That is accomplished with an asymmetrical mechanism to the one employed in the residual feature attention blocks and can be summarized by the following formula:
\begin{equation}
F_{RTA}(\mathrm{\textbf{X}}^{(i)})=\mathrm{\textbf{X}}^{(i)} + H_{TA}(\mathrm{\textbf{X}}^{(i)}_{*})\cdot\mathrm{\textbf{X}}^{(i)}_{*}
\label{residual_temporal_attention}
\end{equation}
where $H_{TA}$ is the temporal attention function and $\mathrm{\textbf{X}}^{(i)}_{*}$ is the product of a stack of two 2D convolutional operations as depicted in Fig. \ref{fig:temporal_attention} with $f_{h}\times f_{w}$ and $T\cdot C$ as filter size and number of features, respectively. Then, as already introduced with the feature attention blocks, the temporal block takes the temporal-wise global spatial information into a feature descriptor by using a global average pooling operation. Finally those statistical descriptors are processed by a stack of 2D convolutional layers with ReLU and sigmoid as activation function, respectively, scaling the $T \cdot C$ channels of the input tensor, as shown in Eq. \ref{residual_temporal_attention}. As for feature attention blocks, the first convolutional layer reduces the number of the last dimension by a factor of $r$, giving the network the possibility to fully capture temporal-wise dependencies from the aggregated output information of the global average pooling operation.
\subsection{Temporal reduction blocks}
The aim of the last block of the main branch is to slowly reduce the number of temporal steps so that the temporal depth eventually reduces to one. Indeed, the output tensor $X^{(i)}_{I}$ of the N residual feature attention blocks has $T$ temporal dimensions that need to be merged. To this end, we further process the incoming tensors with $\lfloor T/(f_{t}-1) \rfloor -1$ temporal reduction blocks. Each one is composed of a residual feature attention block and a 3D convolutional layer without any reflecting padding in the temporal dimension, denoted TR-Conv. So, at each TR-Conv layer we reduce $T$ of $f_{t}-1$. The attention blocks allow the network to learn the best space to decouple image features, "highlighting" more promising features to maintain when reducing the temporal dimension. The output of the last temporal reduction block is a tensor $X^{(i)}_{II}$ with shape $H\times W\times f_{t}\times F$ where the temporal dimension $T$ is reduced to $f_{t}$. The last TR-Conv, before the upsampling
function $H_{\text{UP}|_\text{3D}}$, reduces to one the number of temporal steps and generates $s^{2} \cdot C$ features for the sub-pixel convolutional layer.
\subsection{Training process}
\label{sec:training}
Learning the end-to-end mapping function $H_{\text{RAMS}}$ requires the estimation of model parameters $\Theta$.
That is achieved by minimizing a loss $\mathcal{L}$ between the reconstructed super-resolved images $I^\text{SR}$ and the corresponding ground truth high-resolution images $I^\text{HR}$.

Several loss functions have been proposed and investigated for the SISR problem, such as $L_{\text{1}}$ \cite{lai2017deep, lim2017enhanced, lai2018fast, yu2018wide}, $L_{\text{2}}$ \cite{molini2019deepsum, dong2015image, tai2017memnet, kim1990recursive} and perceptual and adversarial losses \cite{ledig2017photo, johnson2016perceptual}. However, in typical MISR remote-sensing problems, LR images are taken within a certain time window and they could have an undefined spatial misalignment one to each other. So, we must take into account that the super-resolved output of the model $I^\text{SR}$ will be inherently not registered with the target image $I^\text{HR}$. Moreover, since we can have very different conditions among the images part of the same scene, it is important to make the loss function independent from possible intensity biases between the super-resolved $I^\text{SR}$ and the target $I^\text{HR}$. Indeed, if we get a super-resolved image $I^{\text{SR}} = I^{\text{HR}} + \epsilon$, with $\epsilon$ constant and low enough to avoid numerical saturation, we can consider its reconstruction perfect, since it represents the scene with the same level of detail of the ground truth.

With these premises, inspired by the metric proposed in \cite{martens2019super}, we defined $I^\text{SR}_\text{crop}$ as the super-resolved output cropped of $d$ pixels on each border and we consider each possible patch $I^\text{HR}_{u,v},\; u,v\in[0,2d]$ of size $(sH-2d)\times(sW-2d)$ extracted from the ground truth $I^\text{HR}$. We compute the mean biases between the cropped $I^\text{SR}_\text{crop}$ and the patches $I^\text{HR}_{u,v}$ as follows:

\begin{equation}
    \text{b}_{u,v}=\frac{\sum_{i=1}^{sH-2d}\sum_{j=1}^{sW-2d} \big[I^\text{HR}_{u,v}-I^\text{SR}_{u,v}\big](i,j)}{(sH-2d)(sW-2d)} 
\label{loss_bias}
\end{equation}

The loss is then defined as the minimum mean absolute error ($L_{\text{1}}$ loss) between $I^\text{SR}_\text{crop}$ and each possible alignment patch $I^\text{HR}_{u,v}$. We use the MAE instead of the most used MSE since we experimentally find that provides better results for image restoration problems, as proved by the previous works\cite{zhao2016loss,lim2017enhanced,zhang2018image}.

\begin{equation}
    \mathcal{L} = \min_{u,v\in[0,2d]}
    \frac{\|I^\text{HR}_{u,v}-(I^\text{SR}_{u,v}+b_{u,v})\|_1}{(sH-2d)(sW-2d)}
\label{loss}
\end{equation}

\noindent where $\|\cdot\|_1$ represents the $L_{\text{1}}$ norm of a matrix, i.e. the sum of its absolute values.

%----------- 

%------------ Francesco's work
\section{Experiments} \label{experiments}
In this section, we test the proposed methodology in an experimental context, training it on a dataset of real-world satellite images and evaluating its performance in comparison with other approaches, including a state-of-the-art SISR algorithm, to demonstrate the superiority of Multi-image models. We first present the dataset and the preprocessing stages, we define all the parameters used during the experimentation, and then we propose quantitative and qualitative results. We also perform an ablation study to demonstrate the contribution of the global residual branch that implements a temporal attention mechanism. To implement our network, we use the TensorFlow framework. The complete code with a pre-trained version of our model is available online \footnote{https://github.com/EscVM/RAMS}.

\subsection{The Proba-V Dataset}
To train our model, we exploit the dataset released by the Advanced Concept Team of the European Space Agency (ESA) \cite{martens2019super}. This dataset has been specifically conceived for MISR problems, and it is composed of several images taken by the Proba-V satellite \footnote{https://esa.int/Applications/Observing\_the\_Earth/Proba-V.} in the two different spectral bands RED and NIR (near-infrared). Proba-V satellite has been launched by ESA in 2013 and is specifically designed for land covering and vegetation growth monitoring across almost the entire globe. The satellite provides images in two resolutions with different revisit frequency. HR images have a 100m per pixel spatial resolution and are released roughly every five days, while LR images have 300m per pixel resolution and are available almost daily. The characteristics of the Proba-V imagery make it particularly suitable for MISR algorithms since it provides both resolutions natively, allowing for the application of the SR process without the need for artificially degrading and downsampling the HR images.

The dataset has been released for the Proba-V Super Resolution challenge \footnote{https://kelvins.esa.int/proba-v-super-resolution.} and is composed of two main parts: the train part provides both LR and HR images, while the test part LR images, only. In order to verify the effectiveness of our approach, we consider the train part and not the test part, since it has been conceived for the challenge evaluation only and it does not include the ground truths. Thus, we subdivide the train part in training and validation sets. To ease the comparison with previous methods, we use the same validation images used in \cite{molini2019deepsum}. In total, we have 415 scenes for training and 176 for validation for the RED band and 396 for training and 170 for validation for NIR.

Each scene is composed of several LR images (from 9 to 35, depending on the scene) with a dimension of 128x128 pixels and a single HR ground truth with a dimension of 384x384 pixels. The images are encoded as 16-bit png files, even if the actual signal bit-depth is 14 bits. Additionally, each image features a binary mask that distinguishes reliable pixels from unreliable ones (e.g., due to cloud coverage). This information is vital since the images are not taken in the same weather and temporal conditions, but a maximum period of 30 days can be covered in a single scene. For this reason, non-trivial changes in the landscape can occur between different LR images and their HR counterpart and are essential to understand which pixels carry meaningful information and which do not. Trying to infer the value of pixels that are concealed by clouds would mean being able to predict the weather in an unknown time by merely looking at the condition in other unspecified moments. For this reason, it is essential to train the network so that unreliable pixels do not influence the SR process. To assess the quality of each image, we define $c$ as the clearance of the image, i.e. the fraction of reliable pixels in the correspondent binary mask.

\subsection{Data pre-processing}
\label{sec:preprocessing}
Before training the model, we pre-process the dataset with the following steps:
\begin{itemize}
    \item register each LR image using as reference the one with maximum clearance $c$
    \item select the clearest $T$ images from each scene that are above a certain clearance threshold $c_\text{min}$
    \item pre-augment the training dataset with $n_p$ temporal permutations of the LR input images
    \item normalize the images by subtracting the dataset mean intensity value and dividing by the standard deviation
\end{itemize}
Since each LR image is taken at a different time and with some intrinsic spatial misalignment with respect to the others, it is important to resample each pixel value in order to have a coherent reference frame. For each scene of the dataset, we consider as a reference image the one with the maximum clearance $c$. During the registration process, we consider translation as transformation model, which computes the necessary shifts to register each image for both the axes. Masks are taken into consideration during this process in order to avoid bad registration caused by unreliable pixels. The registration is performed in the Fourier domain using normalized cross-correlation as in \cite{padfield2011masked}. After computing the shifts, both LR images and the correspondent masks are shifted accordingly. We use a reflect padding to add pixels to LR images and a constant zero padding for masks. In this way, these extra pixels will be considered unreliable.

For each scene, we must select some LR images in order to match the temporal dimension $T$ of the network. We set a threshold $c_\text{min}= 0.85$ on the clearance for an image to be accepted to avoid using awful images that can worsen the SR performance. The acceptable images are then sorted in order of clearance, and the best $T$ are selected. In the case of a scene with less than $T$ images, we sample randomly from the set of acceptable images until $T$ are reached. If a scene is only composed of clearances under $c_\text{min}$, it is entirely removed from the dataset. This selection process is performed after the registration so that heavily bad registered images are also removed, even if they had an initial clearance above the threshold. Since each scene of the dataset contains at least 9 LR images, we set $T=9$ to fully exploit all the available information for most of the scenes.

One of the characteristics of the Proba-V dataset is that the LR images of a particular scene have no clear temporal order. Therefore, there is no reason to prefer a specific order in the $T$ input images to another. The training dataset is, therefore, pre-augmented by performing $n_p$ random temporal permutations of the selected $T$ input images to help generalization. In this way, we can train the algorithm to identify the best temporal image independently on the position inside the input tensor. We set this permutation parameter to $n_p=7$, reaching a total of 2905 training data-points for RED and 2751 for NIR.

Finally, each image is normalized by subtracting the mean pixel intensity value computed on the entire dataset and dividing by the standard deviation. After the forward pass in the network, all the tensors are then denormalized, and the subsequent evaluations are performed on the 16 bits unsigned integer arrays.

\subsection{Experimental settings}
The scaling factor of the Proba-V dataset is $s=3$. Since we have different scenes for RED and NIR data, we treat the problem for the two bands separately. For this reason, we have $C=1$, since we consider images with a single channel. We set $F=32$ and $f_{h}=f_{w}=f_{t}=3$ as number of filters and kernel size respectively for each convolutional layer. Therefore, the number of temporal reduction blocks is $\lfloor T/(f_{t}-1)\rfloor-1=3$, since each block reduces the temporal dimension of 2. In all the residual attention blocks, we set $r=8$ as the reduction factor. After testing different values with a grid search, we set $N=12$ as the number of residual feature attention blocks in the main branch of the network. We find that decreasing this number causes a loss of performance while increasing it gives a little improvement in the results at the cost of a high increase in the number of parameters. $N=12$ is, therefore, the best compromise between network size and prediction results. In total, our model has slightly less than 1M parameters.

In most of the SR applications present in literature, LR images are obtained from the artificial degradation of the target HR images. In contrast, the real-world nature of the dataset, in which LR images are obtained independently from HR images, causes an unavoidable misalignment between the super-resolved output and the ground truth. To take into account this problem, the authors of the dataset consider a maximum shift of $\pm3$ pixels on each axis between $I^{\text{SR}}$ and target $I^{\text{HR}}$, computed on the basis of the geolocation accuracy of the Proba-V satellite \cite{martens2019super}. When computing the loss function presented in Sec. \ref{sec:training}, we can therefore set $d = 3$. Besides, since the Proba-V dataset also provides binary mask that marks with one reliable pixel and with 0 unreliable (e.g., concealed by clouds) ones, we adapt the loss function to use this information to refine the training process.  During the loss computation, we want pixels marked as unreliable in the target binary mask $M^\text{HR}$ not to influence the loss computation. Practically, we can simply multiply the cropped super-resolved image $I^\text{SR}_\text{crop}$, and the HR patch $I^\text{HR}_{u,v}$ by the correspondent cropped mask $M^\text{HR}_{u,v}$ and average all the quantities over the number of clear pixels. The bias computation is therefore adapted from Eq. \ref{loss_bias} as:

\begin{equation}
    \text{b}_{u,v}=\frac{\sum_{i,j}\big[I^\text{HR}_{u,v}\cdot M^\text{HR}_{u,v}
    -I^\text{SR}_{u,v}\cdot M^\text{HR}_{u,v}\big](i,j)}{\|M^\text{HR}_{u,v}\|_1}
\label{loss_bias_masked}
\end{equation}

\noindent where $\|\cdot\|_1$ represents the $L_\text{1}$ norm of a matrix, i.e. the sum of its  absolute values. In the same way, the loss is adapted from Eq. \ref{loss} as:

\begin{equation}
    \mathcal{L} = \min_{u,v\in[0,6]}\frac{\|I^\text{HR}_{u,v}\cdot M^\text{HR}_{u,v} -(I^\text{SR}_{u,v}\cdot M^\text{HR}_{u,v}+b_{u,v})\|_1}{\|M^\text{HR}_{u,v}\|_1}
\label{loss_masked}
\end{equation}

To train the model, we extract from each LR image 16 patches with a size of $32\times32$ pixels and the corresponding HR and masks patches with a size of $96\times96$. We further check every single patch and remove those that have a target mask $M^\text{HR}$ with less than 0.85 clearance. The total number of training data points obtained is 41678 for RED and 40173 for NIR. During the training process, we further perform data augmentation with random rotations of $90^\circ$, $180^\circ$ and $270^\circ$ and random horizontal flips.

We set the batch size to $32$. Therefore, during training, we have an input tensor with shape $32\times32\times32\times9\times1$ and an output tensor with shape $32\times96\times96\times1$. We optimize the loss function with Adam algorithm \cite{kingma2014adam} with default parameters $\beta_1=0.9$, $\beta_2=0.999$ and $\epsilon=10^{-7}$. We set an initial learning rate 
$\eta_\text{i}=5\times10^{-4}$ and we reduce it with a linear decay down to $\eta_\text{f}=5\times10^{-7}$. We train two different networks for RED and NIR spectral bands on a workstation with an Nvidia RTX 2080Ti GPU with 11GB of memory and 64GB of DDR4 SDRAM. We use the TensorFlow 2.0 framework with CUDA 10. In total, we train the models for 100 training epochs for about 16 hours.

\subsection{Quantitative results}
To evaluate the obtained results, we need to use a slightly modified version of PSNR and SSIM \cite{wang2004image} criteria to take into consideration all the aspects we considered in the previous section to obtain a proper loss function. Martens et al. \cite{martens2019super} propose a corrected version of the PSNR, called cPSNS, that is obtained from a corrected mean squared error (cMSE). The computation of the cMSE is performed in the same way as we did for the loss in Eq. \ref{loss_masked}: it is the minimum MSE between $I^\text{SR}_\text{crop} + b_{u,v}$ and the HR patches $I^{\text{HR}}_{u,v}$:

\begin{equation}
    \text{cMSE}=\min_{u,v\in[0,6]}\;\underset{\text{clear}}{\text{MSE}}\big(I^\text{HR}_{u,v},
    I^\text{SR}_\text{crop} + b_{u,v}\big)
\end{equation}

\noindent where $\underset{\text{clear}}{\text{MSE}}$ represents the mean squared error computed only on pixels marked as clear in the binary mask $M^\text{HR}_{u,v}$. Again, we can simply multiply the matrices by the mask to make unreliable pixels irrelevant:

\begin{equation}
    \underset{\text{clear}}{\text{MSE}}=\frac{\|I^\text{HR}_{u,v}\cdot M^\text{HR}_{u,v}
    -(I^\text{SR}_{u,v}\cdot M^\text{HR}_{u,v}+b_{u,v})\|_2^2}{\|M^\text{HR}_{u,v}\|_1}
\end{equation}

 \noindent where $\|\cdot\|_2$ represents the Frobenius ($L_\text{2}$) norm of a matrix, i.e. the square root of the sum of its squared values. We can then compute the cPSNR as:

\begin{equation}
\begin{split}
    \text{cPSNR} &= 10\log_{10} \frac{(2^{16}-1)^2}{\text{cMSE}} \\
                 &= \max_{u,v\in[0,6]}10\log_{10}\;\frac{(2^{16}-1)^2}
                 {\underset{\text{clear}}{\text{MSE}}(I^\text{HR}_{u,v},
                 I^\text{SR}_\text{crop} + b_{u,v})}
\end{split}
\end{equation}
where $2^{16}-1$ is the maximum pixel intensity for an image encoded on 16 bits.

In the same way, we can define a corrected version of the SSIM metric: cSSIM is the maximum SSIM between $I^\text{SR}_\text{crop} + b_{u,v}$ and the HR patches $I^{\text{HR}}_{u,v}$, again multiplied for the mask $M^{\text{HR}}_{u,v}$.

\begin{equation}
    \text{cSSIM} = \max_{u,v\in[0,6]}\text{SSIM}\big(I^\text{HR}_{u,v}\cdot M^\text{HR}_{u,v},
    I^\text{SR}_\text{crop}\cdot M^\text{HR}_{u,v} + b_{u,v}\big)
\end{equation}

\subsubsection{Temporal self-ensemble (RAMS+)}
\begin{figure}[h]
\centering
\begin{tabular}{c}
\includegraphics[width=\columnwidth]{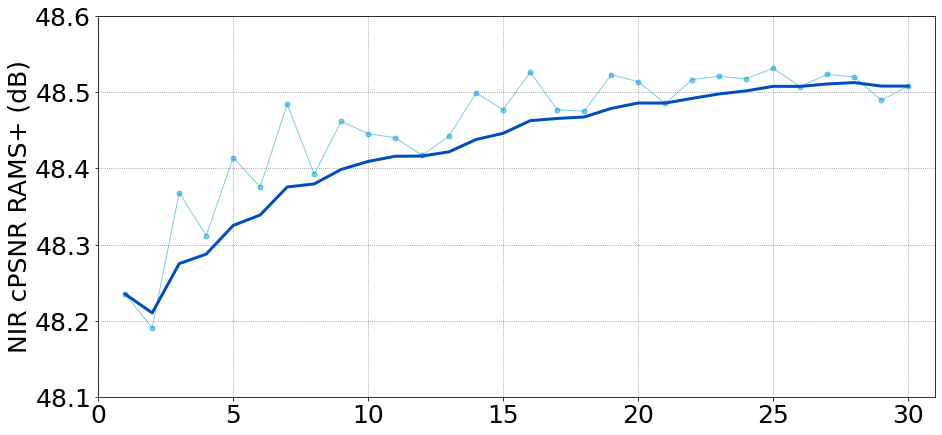}\\
\includegraphics[width=\columnwidth]{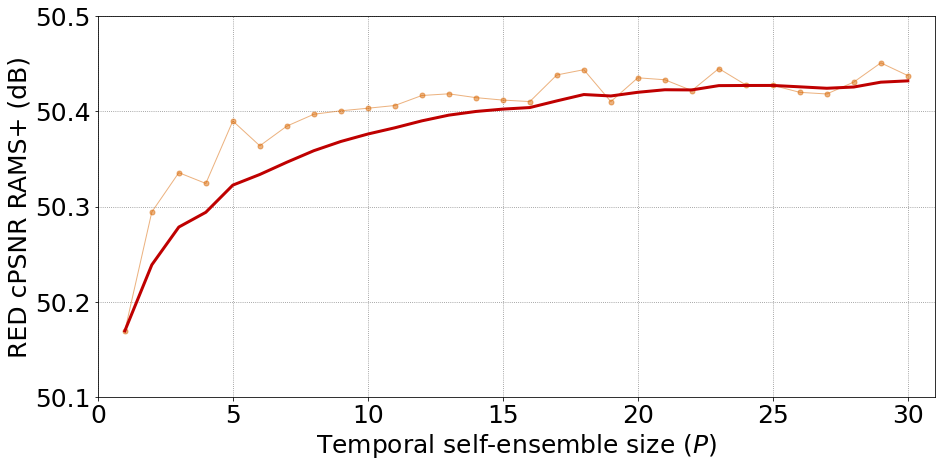}
\end{tabular}
\caption{Results with a temporal self-ensemble of size $P$. The highlighted curves represent an exponential moving average of the results to clearly show the trend. The values for $P=1$ are equivalent to RAMS.}
\label{fig:RAMS+}
\end{figure}
As in Sec. \ref{sec:preprocessing}, during the training process images are augmented with random permutation in the temporal axis. For this reason, it is possible to maximize the performance of the model, by adopting a self-ensemble mechanism during inference, similarly to what done in previous super-resolution works \cite{lim2017enhanced,timofte2016seven,zhang2018image}. For each input scene, we consider a certain number $P$ of random permutations on the temporal axis and we denote as $\big\{I^\text{LR}_{[1,T],\,0}\;,\cdots,\; I^\text{LR}_{[1,T],\,P}\big\}$ the resulting set of inputs. The output of the inference process is therefore the average of the predictions on the whole set. We call this methodology RAMS+$_{P}$, where $P$ is the number of random permutations performed:

\begin{equation}
I^{\text{SR}}=\frac{1}{P}\sum_{i=1}^{P} H_\text{RAMS}\big(I^{\text{LR}}_{[1,T],\,i}\big)
\end{equation}

Fig. \ref{fig:RAMS+} shows cPSNR results on the testing dataset for a different number of permutated predictions. The trend clearly shows how increasing $P$ results in better performance on both the spectral bands, with an effect that tends to saturate for $P\geq25$. For the following evaluation, we select $P=20$ to present the results for RAMS+. It is worth noting that, even if this method allows to increase the performance of the network sharply, inference time grows linearly with $P$, with RAMS+$_{20}$ taking roughly 20 times as long as RAMS. Another aspect to highlight is that the permutations are performed randomly, so different results can be achieved even with the same value of $P$.

\subsubsection{Comparison with state-of-the-art methods}
\begin{figure*}[h]
\centering
\begin{tabular}{cccc}
 \includegraphics[width=0.46\columnwidth]{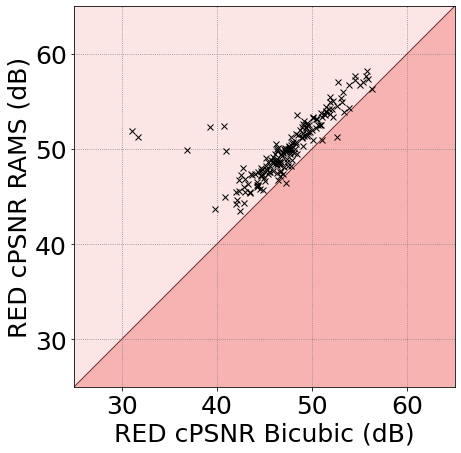}
&\includegraphics[width=0.46\columnwidth]{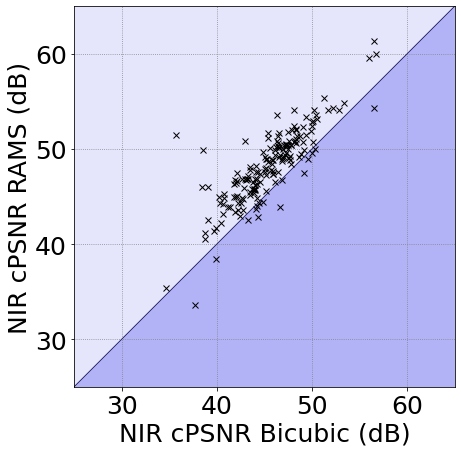}
&\includegraphics[width=0.46\columnwidth]{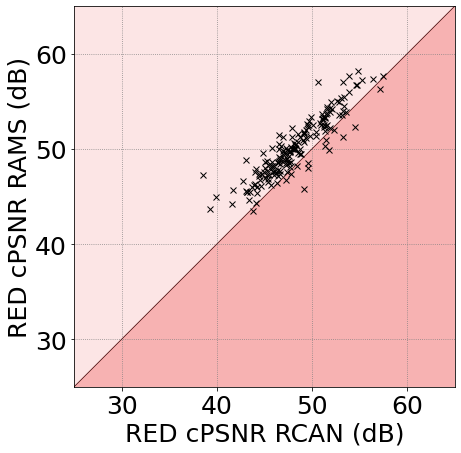}
&\includegraphics[width=0.46\columnwidth]{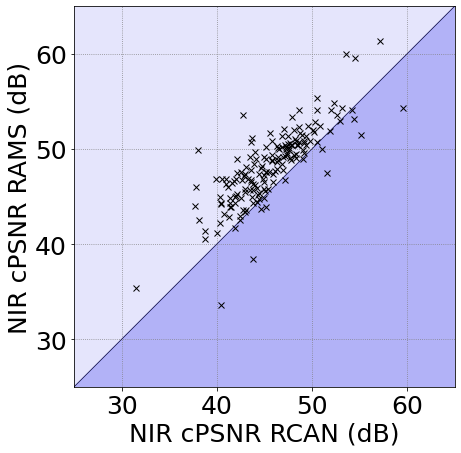}
\end{tabular}
\caption{cPSNR comparison between RAMS and bicubic inteprolation and RAMS and RCAN(SISR) on the validation set. Each data point represents a scene of the dataset: when a cross is above the line, the correspondent scene is reconstructed better by RAMS.}
\label{fig:RAMS_comparison}
\end{figure*}

\begin{table*}[]
\centering
\begin{tabular}{l|cc|cc}
\multicolumn{1}{c|}{Band}   & \multicolumn{2}{c|}{NIR} & \multicolumn{2}{c}{RED} \\ \hline
\multicolumn{1}{c|}{Metric} & cPSNR       & cSSIM      & cPSNR      & cSSIM      \\ \hline\hline
Bicubic                     & 45.12       & 0.9767     & 47.63      & 0.9846     \\
IBP\cite{irani1991improving}& 45.96       & 0.9796     & 48.21      & 0.9865     \\
BTV\cite{farsiu2004fast}    & 45.93       & 0.9794     & 48.12      & 0.9861     \\
RCAN\cite{zhang2018image}   & 45.66       & 0.9798     & 48.22      & 0.9870     \\
VSR-DUF\cite{jo2018deep}    & 47.20       & 0.9850     & 49.59      & 0.9902     \\
HighRes-net\cite{deudon2020highres} & 47.55       & 0.9855     & 49.75      & 0.9904     \\
DeepSUM\cite{molini2019deepsum}    & 47.84       & 0.9858     & 50.00      & 0.9908     \\
DeepSUM++\cite{molini2020deepsum++} & 47.93       & 0.9862     & 50.08      & 0.9912     \\
\textbf{RAMS (ours)}        &\textbf{48.23} &\textbf{0.9875} &\textbf{50.17} &\textbf{0.9913} \\
\textbf{RAMS+$_{20}$ (ours)}&\textbf{48.51} &\textbf{0.9880} &\textbf{50.44} &\textbf{0.9917}
\end{tabular}
\caption{Average cPSNR (dB) and cSSIM over the validation dataset for different methods.}
\label{tab:results}
\end{table*}

Tab. \ref{tab:results} shows the comparison of cPSNR and cSSIM metrics with several methods on the validation set. We consider as the baseline the bicubic interpolation of the best image of the scene selected considering the clearance, i.e., the number of clear pixels as marked by the binary masks.\\
IBP\cite{irani1991improving} and BTV\cite{farsiu2004fast} methods are tested with the same methodology presented in Molini et al. \cite{molini2019deepsum}. They achieve slightly better results than the baseline with both the metrics.\\
RCAN \cite{zhang2018image} is currently one of the Single-image Super-resolution state-of-the-art networks. We trained from scratch two models, one for each spectral band, setting $G=5$ and $B=5$, as the number of residual groups and residual channel attention blocks respectively, for a total of about 2 million parameters. We train the two models from scratch on the Proba-V dataset, selecting the best image per scene as input. RCAN shows better performance with respect to classical methods but is far beyond the other MISR networks, showing how the additional information coming from both spatial and temporal correlations is vital to boost the super-resolution process.

VSR-DUF\cite{jo2018deep} has been developed to upsample video signals using a temporal window of several frames. We train two models from scratch, one for each spectral bands, using 9 LR images as input as in our methodology. The authors consider three different architectures depending on the number of convolutional layers and find better results, increasing the depth of the model. We select the baseline 16 layers deep architecture, that already has more than double parameters with respect to RAMS, with the same number of input images.

HighRes-net\cite{deudon2020highres} algorithm got the second place in the Proba-V challenge and featured a single network for both spectral bands that recursively reduce the temporal dimension to fuse the input LR images. We train the model on our training dataset with default architectures. Since the authors designed the architecture to have an input temporal dimension multiple of 2, we set it to 16, as it is closest to 9.

DeepSUM \cite{molini2019deepsum} is the algorithm winner of the original Proba-V challenge, and the authors have further developed it with DeepSUM++\cite{molini2020deepsum++}. We train our RAMS on the same training dataset as these two works.
\begin{figure*}[h!]
\centering
\begin{subfigure}{.33\textwidth}
  \centering
  \includegraphics[width=\linewidth]{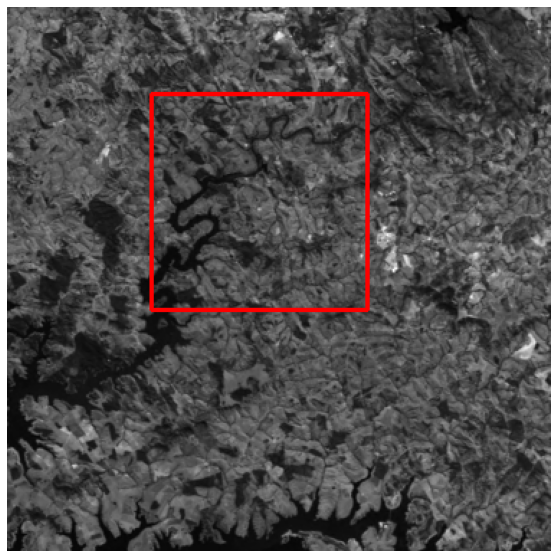}
  \caption*{}
\end{subfigure}
\begin{subfigure}{.66\textwidth}
\vspace*{3pt}
    \centering % <-- added
    \begin{subfigure}{0.21\linewidth}
      \centering
      \includegraphics[width=\linewidth]{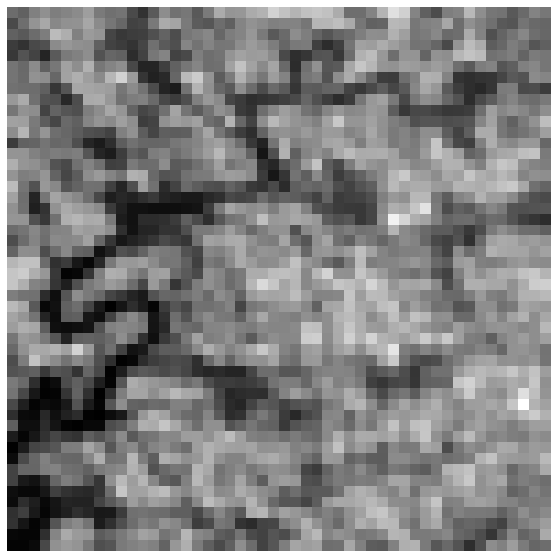}
      \vspace*{-17pt}\caption*{\scalebox{0.85}{LR}}
      \vspace*{-8pt}\caption*{\scalebox{0.85}{(cPSNR / cSSIM)}}
    \end{subfigure}\hfil % <-- added
    \begin{subfigure}{0.21\linewidth}
      \includegraphics[width=\linewidth]{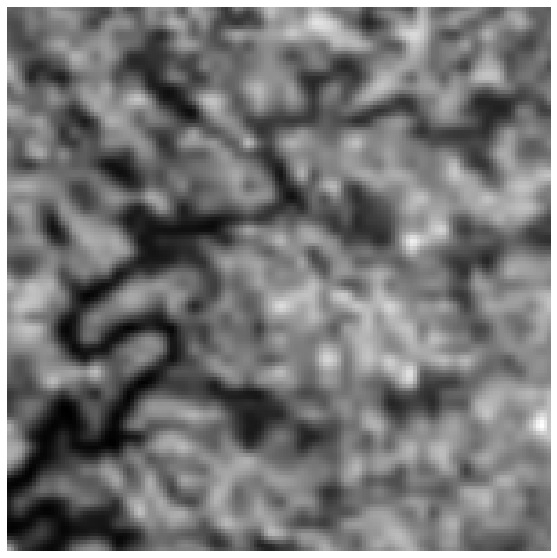}
      \vspace*{-17pt}\caption*{\scalebox{0.85}{Bicubic}
      \vspace*{-8pt}}\caption*{\scalebox{0.85}{(48.30 / 0.9857)}}
    \end{subfigure}\hfil % <-- added
    \begin{subfigure}{0.21\linewidth}
      \includegraphics[width=\linewidth]{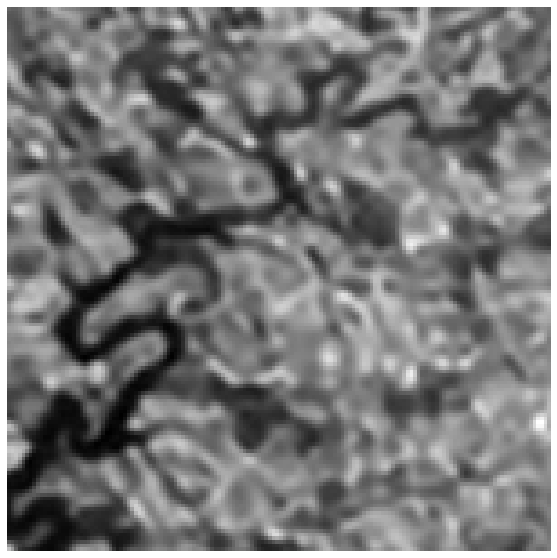}
      \vspace*{-17pt}\caption*{\scalebox{0.85}{RCAN\cite{zhang2018image}}}
      \vspace*{-8pt}\caption*{\scalebox{0.85}{(49.18 / 0.9887)}}
    \end{subfigure}\hfil
    \begin{subfigure}{0.21\linewidth}
      \includegraphics[width=\linewidth]{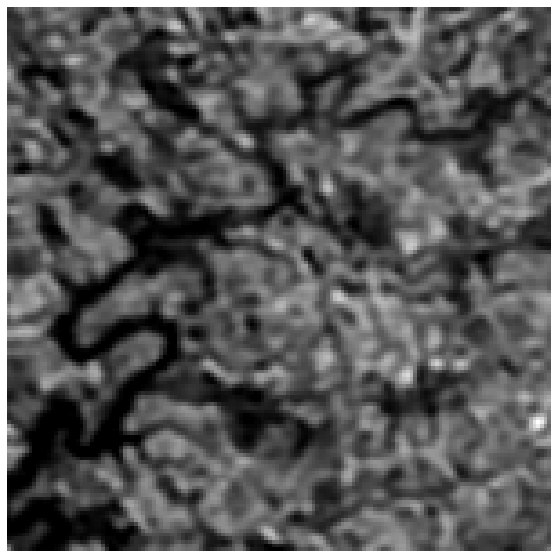}
      \vspace*{-17pt}\caption*{\scalebox{0.85}{VSR-DUF\cite{jo2018deep}}}
      \vspace*{-8pt}\caption*{\scalebox{0.85}{(50.29 / 0.9909)}}
    \end{subfigure}
\vskip 3pt
    \begin{subfigure}{0.21\linewidth}
      \includegraphics[width=\linewidth]{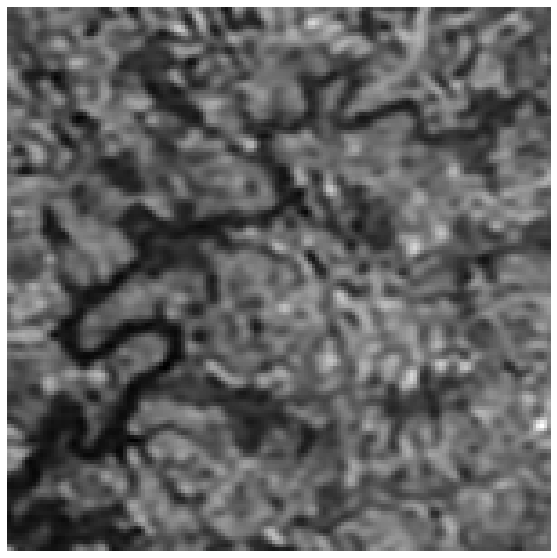}
      \vspace*{-17pt}\caption*{\scalebox{0.85}{DeepSUM\cite{molini2019deepsum}}}
      \vspace*{-8pt}\caption*{\scalebox{0.85}{(50.73 / 0.9917)}}
    \end{subfigure}\hfil % <-- added
    \begin{subfigure}{0.21\linewidth}
      \includegraphics[width=\linewidth]{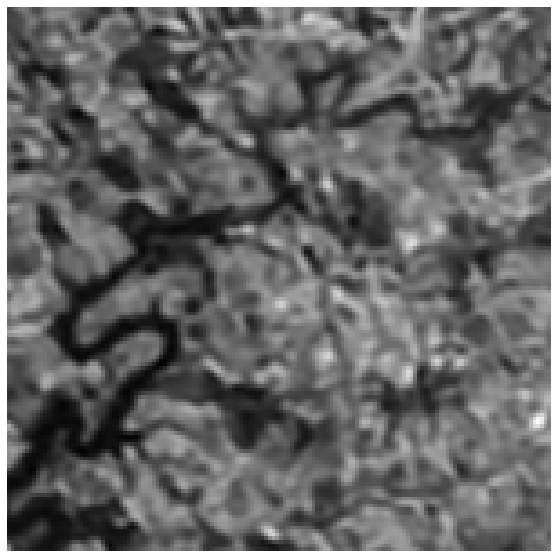}
      \vspace*{-17pt}\caption*{\scalebox{0.85}{\textbf{RAMS}}}
      \vspace*{-8pt}\caption*{\scalebox{0.85}{(51.53 / 0.9930)}}
    \end{subfigure}\hfil % <-- added
    \begin{subfigure}{0.21\linewidth}
      \includegraphics[width=\linewidth]{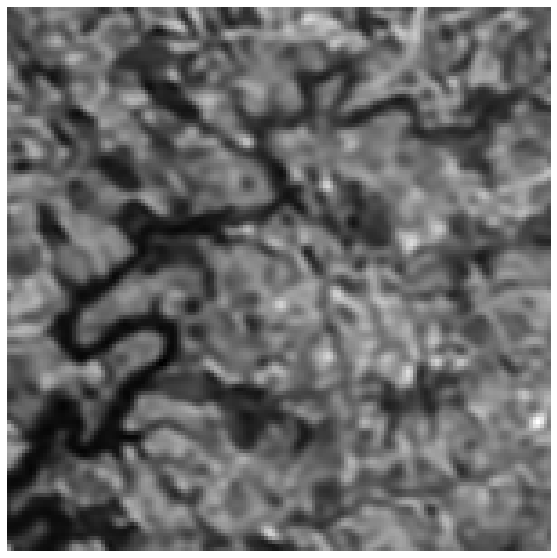}
      \vspace*{-17pt}\caption*{\scalebox{0.85}{\textbf{RAMS+$_{20}$}}}
      \vspace*{-8pt}\caption*{\scalebox{0.85}{(51.64 / 0.9932)}}
    \end{subfigure}\hfil
    \begin{subfigure}{0.21\linewidth}
      \includegraphics[width=\linewidth]{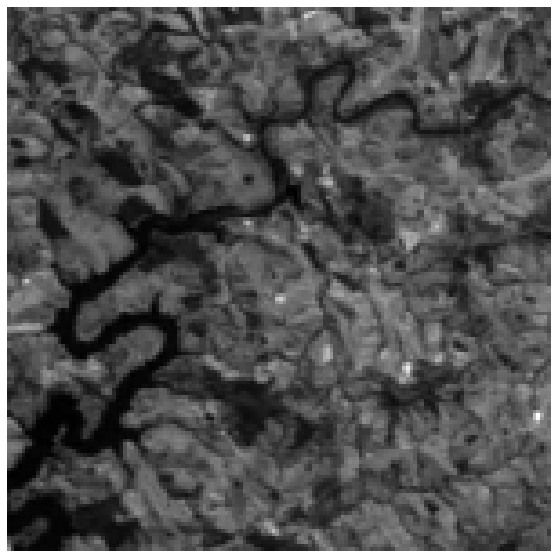}
      \vspace*{-17pt}\caption*{\scalebox{0.85}{HR}}
      \vspace*{-8pt}\caption*{\scalebox{0.85}{(cPSNR / cSSIM)}}
    \end{subfigure}
\end{subfigure}
\caption{Qualitative comparison between different methods on RED imgset0302.}
\label{fig:results_RED}
\end{figure*}

\begin{figure*}[h!]
\centering
\begin{subfigure}{.33\textwidth}
  \centering
  \includegraphics[width=\linewidth]{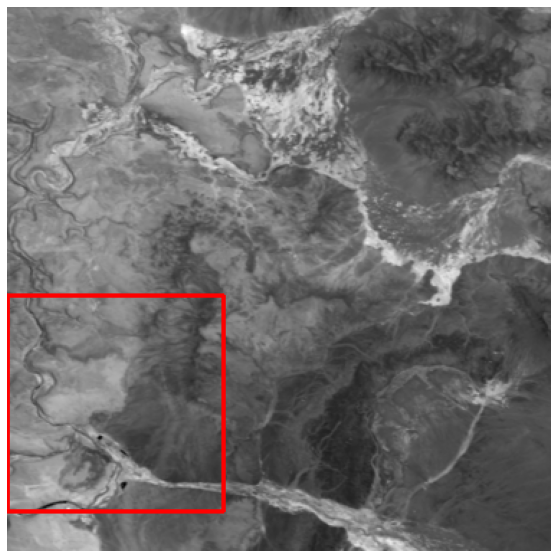}
  \caption*{}
\end{subfigure}
\begin{subfigure}{.66\textwidth}
\vspace*{3pt}
    \centering % <-- added
    \begin{subfigure}{0.21\linewidth}
      \centering
      \includegraphics[width=\linewidth]{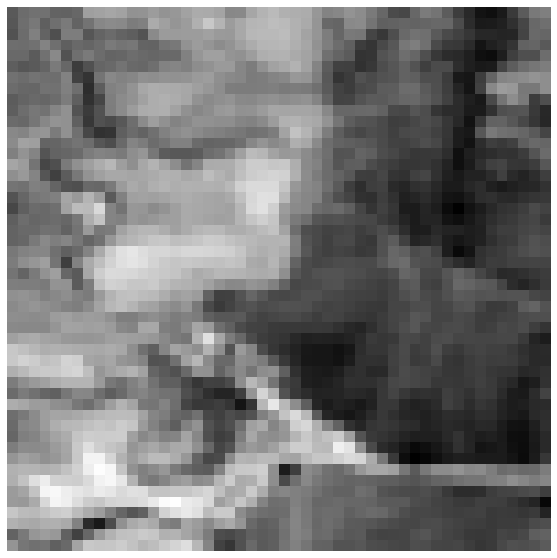}
      \vspace*{-17pt}\caption*{\scalebox{0.85}{LR}}
      \vspace*{-8pt}\caption*{\scalebox{0.85}{(cPSNR / cSSIM)}}
    \end{subfigure}\hfil % <-- added
    \begin{subfigure}{0.21\linewidth}
      \includegraphics[width=\linewidth]{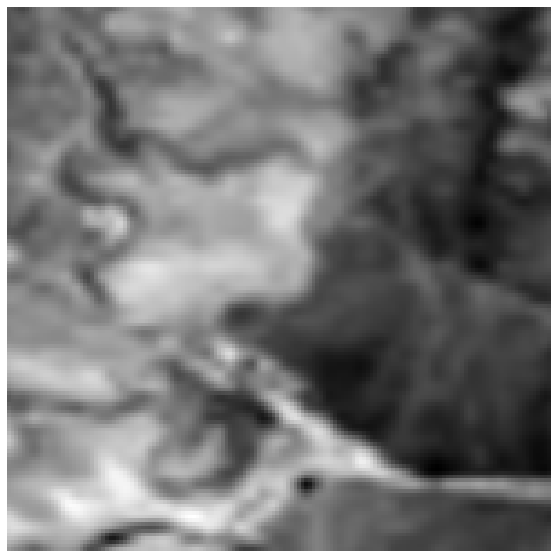}
      \vspace*{-17pt}\caption*{\scalebox{0.85}{Bicubic}
      \vspace*{-8pt}}\caption*{\scalebox{0.85}{(44.09 / 0.9758)}}
    \end{subfigure}\hfil % <-- added
    \begin{subfigure}{0.21\linewidth}
      \includegraphics[width=\linewidth]{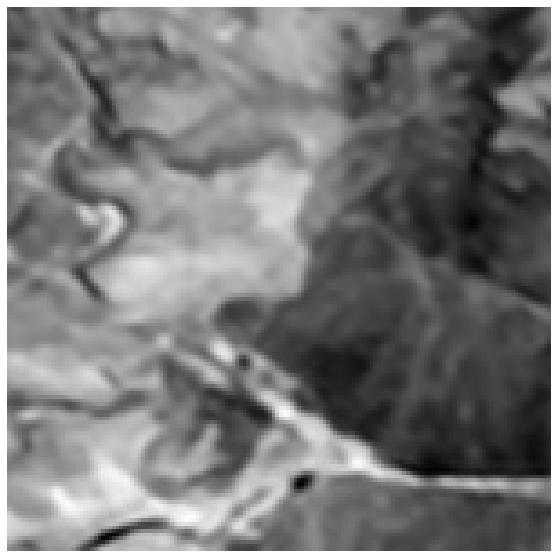}
      \vspace*{-17pt}\caption*{\scalebox{0.85}{RCAN\cite{zhang2018image}}}
      \vspace*{-8pt}\caption*{\scalebox{0.85}{(44.81 / 0.9830)}}
    \end{subfigure}\hfil
    \begin{subfigure}{0.21\linewidth}
      \includegraphics[width=\linewidth]{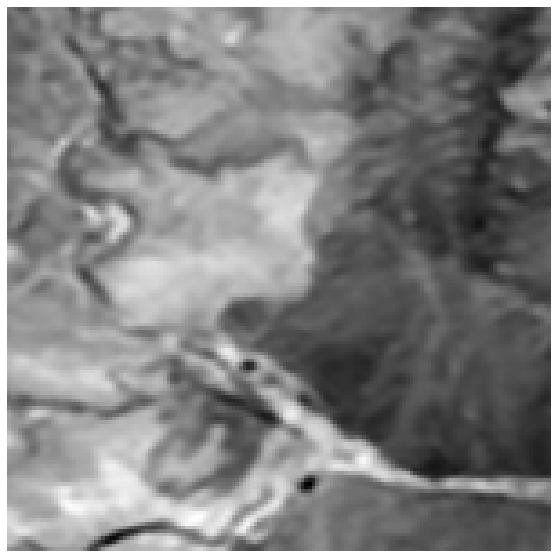}
      \vspace*{-17pt}\caption*{\scalebox{0.85}{VSR-DUF\cite{jo2018deep}}}
      \vspace*{-8pt}\caption*{\scalebox{0.85}{(45.94 / 0.9857)}}
    \end{subfigure}
\vskip 3pt
    \begin{subfigure}{0.21\linewidth}
      \includegraphics[width=\linewidth]{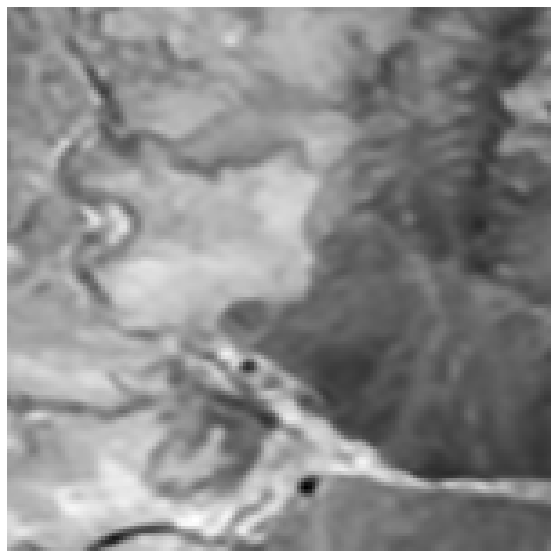}
      \vspace*{-17pt}\caption*{\scalebox{0.85}{DeepSUM\cite{molini2019deepsum}}}
      \vspace*{-8pt}\caption*{\scalebox{0.85}{(47.73 / 0.9887)}}
    \end{subfigure}\hfil % <-- added
    \begin{subfigure}{0.21\linewidth}
      \includegraphics[width=\linewidth]{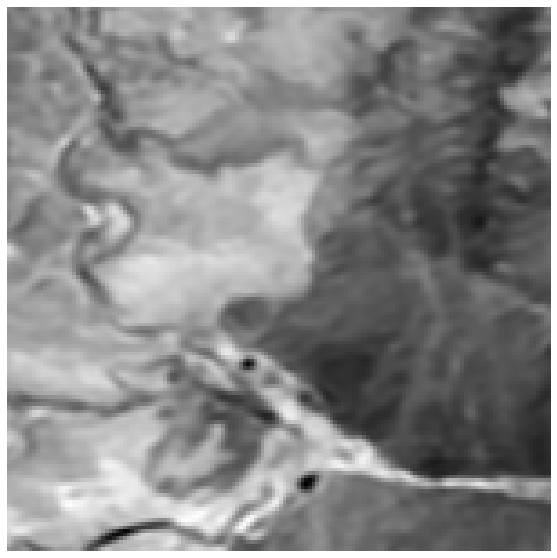}
      \vspace*{-17pt}\caption*{\scalebox{0.85}{\textbf{RAMS}}}
      \vspace*{-8pt}\caption*{\scalebox{0.85}{(48.19 / 0.9899)}}
    \end{subfigure}\hfil % <-- added
    \begin{subfigure}{0.21\linewidth}
      \includegraphics[width=\linewidth]{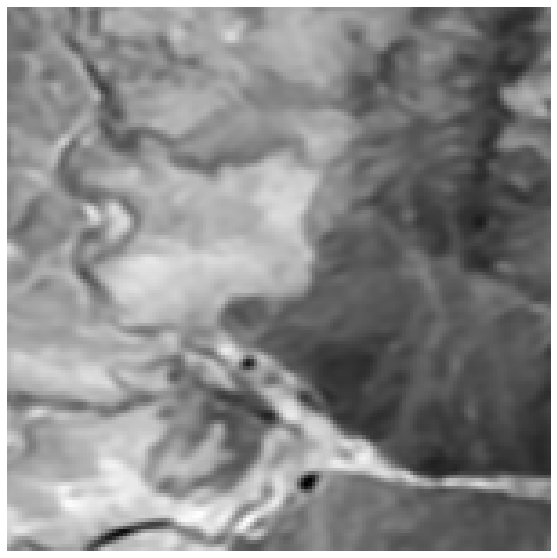}
      \vspace*{-17pt}\caption*{\scalebox{0.85}{\textbf{RAMS+$_{20}$}}}
      \vspace*{-8pt}\caption*{\scalebox{0.85}{(48.92 / 0.9909)}}
    \end{subfigure}\hfil
    \begin{subfigure}{0.21\linewidth}
      \includegraphics[width=\linewidth]{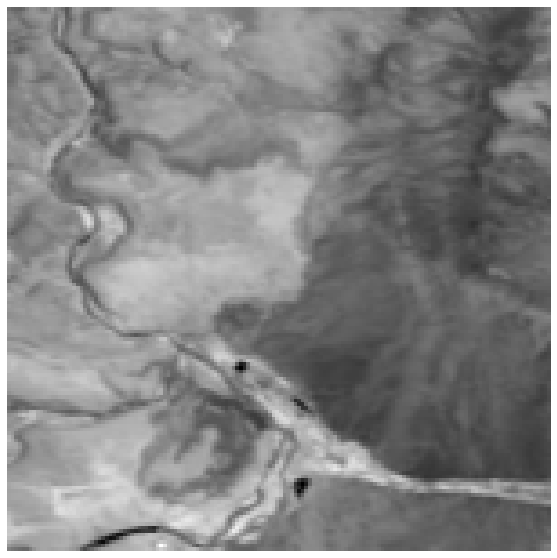}
      \vspace*{-17pt}\caption*{\scalebox{0.85}{HR}}
      \vspace*{-8pt}\caption*{\scalebox{0.85}{(cPSNR / cSSIM)}}
    \end{subfigure}
\end{subfigure}
\caption{Qualitative comparison between different methods on NIR imgset0596.}
\label{fig:results_NIR}
\end{figure*}
Results clearly show how the proposed methodology can obtain the best results with the two metrics on both the spectral bands and thus represents the current state-of-the-art for Multi-image Super-resolution for remote sensing applications. Using temporal self-ensemble, RAMS+ is able to achieve even higher performance. We show the value for RAMS+, setting $P=20$ as the size of the ensemble, which is the value at which we experimentally find that the resulting gain starts to saturate. However, further increasing the ensemble size can result in even better performance, though at the cost of a significant inference speed drop.

It is worth mentioning that our methodology reaches a result of 0.9336790819983855 on the test set of the Proba-V challenge as provided by the official site and places at the top of the leaderboard available after the end of the official challenge\footnote{https://kelvins.esa.int/proba-v-super-resolution-post-mortem/leaderboard.}. This score is computed as the mean ratio between the cPSNR values of the challenge baseline on each testing scene, and the correspondent submitted cPSNR for both the spectral bands. This result has been obtained by retraining the two networks with both training and validation datasets together.

Fig. \ref{fig:RAMS_comparison} shows a direct comparison between the cPSNR results of RAMS and the bicubic interpolation baseline and RCAN (SISR state-of-the-art). Each cross represents a scene of the validation dataset of the corresponding spectral band. The graphs on the left show how our method strongly beats the bicubic upsampling on almost all the scenes, 98\% for RED and 91\% for NIR. That is coherent with a general worse behavior of all the methods on the NIR images, probably due to an intrinsic worse information quality of the NIR dataset. The graphs on the right show, on the other hand, the enormous potential of MISR with respect to SISR methods. It can be observed how again RAMS outperforms RCAN an almost all the scenes, with results only slightly worse than to bicubic interpolation, 92\% for RED, and 91\% for NIR. That is reasonable since RCAN results are someway in the middle between bicubic and RAMS. 

\subsubsection{Importance of the residual temporal attention branch}

As a final analysis, we perform an ablation study to demonstrate the importance of the global residual branch that implements a temporal attention mechanism. We train two alternative networks, one for each spectral band, that have the same architecture of RAMS, except that we delete the residual temporal attention (RTA) branch. These reduced networks are trained from scratch independently from the complete ones. Tab. \ref{tab:ablation} shows a significant drop in the results obtained without the global residual branch and demonstrates the importance of selecting the best temporal views to ease the super-resolution process of the main branch. We find this difference particularly relevant for the RED band, since the training repeatedly failed without the RTA branch, with a diverging behavior after some epochs. The result reported in the table is computed with the last valuable parameters before the divergence starts.

\begin{table}[]
\centering
\begin{tabular}{c|cc|cc}
     & \multicolumn{2}{c|}{without RTA} & \multicolumn{2}{c}{\textbf{with RTA}} \\ \hline
     & cPSNR       & cSSIM       & cPSNR      & cSSIM      \\ \hline\hline
NIR  & 47.96\;\;   & 0.9869\;\;  & 48.23      & 0.9875     \\
RED  & 47.98*      & 0.9863*     & 50.17      & 0.9913     \\
\end{tabular}
\caption{RAMS results with and without RTA (residual temporal attention) branch. Values for RED without RTA are computed with the last valuable parameters before training diverges.}
\label{tab:ablation}
\end{table}

\subsection{Qualitative results}
A visual comparison between some of the methods taken in the exam is shown in Fig. \ref{fig:results_RED} and \ref{fig:results_NIR} for a RED and NIR image respectively. We provide a zoomed patch of the best LR input image of the scene, its bicubic interpolation, and the inference output of RCAN, VSR-DUF, DeepSUM, RAMS and RAMS+$_{20}$, together with the target HR ground truth. cPSNR and cSSIM scores for the image under analysis are also provided. From this comparison, MISR methods clearly show a better performance with respect to bicubic and SISR (RCAN). However, it is not trivial to understand which method is the better among MISR algorithms with a visual inspection of the results, only. As found by Ledig et al. \cite{ledig2017photo}, the task of achieving pleasant-looking results is a different optimization problem from maximizing the fidelity of the reconstructed information. Therefore, results with high content-related metrics as PSNR and SSIM frequently appear less photo-realistic to a human eye. However, in the context of remote sensing, the fidelity of the pixels content is vital to ensure that the super-resolved image are meaningful, thus the quality of results should be inferred by using content-related metrics, rather than by visual inspection.

%----------- 

%----------- All's Work
\section{Conclusion} \label{conclusion}
In this paper, we proposed a novel representation learning model to super-resolve remotely sensed multi-temporal LR images by exploiting concurrently spatial and temporal correlations. We introduced a feature and temporal attention mechanisms with 3D convolutions that, coupled with nestled residual connections, let the network focus on high-frequency components, flow redundant low-frequency information and transcend the local region of convolutional operations. Extensive experiments on the open-source Proba-V MISR dataset, either with single image and multi-image SR methodologies, demonstrated the effectiveness of our proposed methodology. In both NIR and RED spectral bands, our efficient and straightforward solution achieved considerably better results than other literature methodologies obtaining 48.51 dB and 50.44 dB of cPSNR, respectively for the two channels. That is further proved by the score of the official post-mortem Prova-V challenge where RAMS claimed the first place in the leaderboard. Future work may investigate the performance of the RAMS architecture on hyperspectral remote sensing imaging.
%----------- 

\section*{acknowledgments}
This work has been developed with the contribution of the Politecnico di Torino Interdepartmental Centre for Service Robotics PIC4SeR (https://pic4ser.polito.it) and SmartData@Polito (\url{https://smartdata.polito.it}).

\section*{Author Contributions}
Conceptualization, V.M., M.C.; Methodology, V.M.; Software, F.S. and V.M.; Validation, F.S. and V.M.; Data curation, F.S. and V.M.; Writing-original draft, F.S., V.M., and
A.K.; Writing-review and editing, F.S., V.M., A.K. and M.C.; Project administration, V.M. and M.C.; Funding acquisition, M.C. All authors have read and agreed to the published version of the manuscript.

% Can use something like this to put references on a page
% by themselves when using endfloat and the captionsoff option.

\bibliographystyle{IEEEtran}
\bibliography{mainArXiv}

\vfill\break

\begin{IEEEbiography}[{\includegraphics[width=1in,height=1.25in,clip,keepaspectratio]{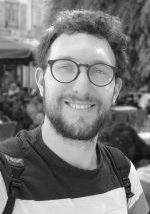}}]{Francesco Salvetti} is currently a Ph.D. student in Electrical, Electronics and Communications Engineering in collaboration with the two interdepartmental centers PIC4SeR (\url{https://pic4ser.polito.it/}) and Smart Data (\url{https://smartdata.polito.it/}) at Politecnico di Torino, Italy. He received his Bachelor's Degree in Electronic Engineering§ in 2017 and his Master’s Degree in Mechatronics Engineering in 2019 at Politecnico di Torino. He is currently working on Machine Learning applied to Computer Vision and Image Processing in robotics applications.
\end{IEEEbiography}

\begin{IEEEbiography}[{\includegraphics[width=1in,height=1.25in,clip,keepaspectratio]{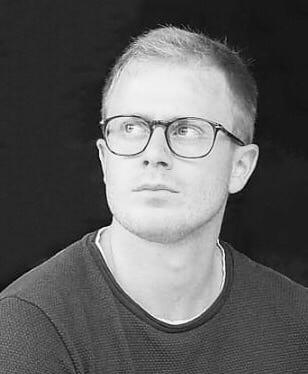}}]{Vittorio Mazzia} is a Ph.D. student in Electrical, Electronics and Communications Engineering working with the two Interdepartmental Centres PIC4SeR (\url{https://pic4ser.polito.it/}) and SmartData (\url{https://smartdata.polito.it/}). He received a master's degree in Mechatronics Engineering from the Politecnico di Torino, presenting a thesis entitled "Use of deep learning for automatic low-cost detection of cracks in tunnels," developed in collaboration with the California State University. His current research interests involve deep learning applied to different tasks of computer vision, autonomous navigation for service robotics, and reinforcement learning. Moreover, making use of neural compute devices (like Jetson Xavier, Jetson Nano, Movidius Neural Stick) for hardware acceleration, he is currently working on machine learning algorithms and their embedded implementation for AI at the edge. 
\end{IEEEbiography}

\begin{IEEEbiography}[{\includegraphics[width=1in,height=1.25in,clip,keepaspectratio]{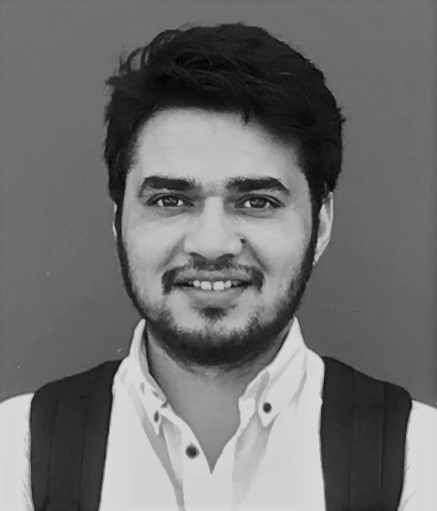}}]{Aleem Khaliq} is a Ph.D. student in Electrical, Electronics and Communications Engineering program at Politecnico di Torino, Italy. He received his Bachelor and Master degree in Electronic Engineering from International Islamic University Islamabad (IIUI), Pakistan, in 2008 
and 2012, respectively. His research interests include remote sensing data analysis, precision agriculture, applications of machine learning in remote sensing, and image processing.  
From 2010 to date, he has been working as a Lab Engineer (currently on study leave) in the Department of Electrical Engineering, Faculty of engineering \& Technology, IIUI. His doctoral studies are funded by Higher Education Commission of Pakistan. Moreover, he is an active member of Polito Inter-departmental Center for Service Robotics (PIC4SeR, \url{https://pic4ser.polito.it/} 
\end{IEEEbiography}

\begin{IEEEbiography}[{\includegraphics[width=1in,height=1.25in,clip,keepaspectratio]{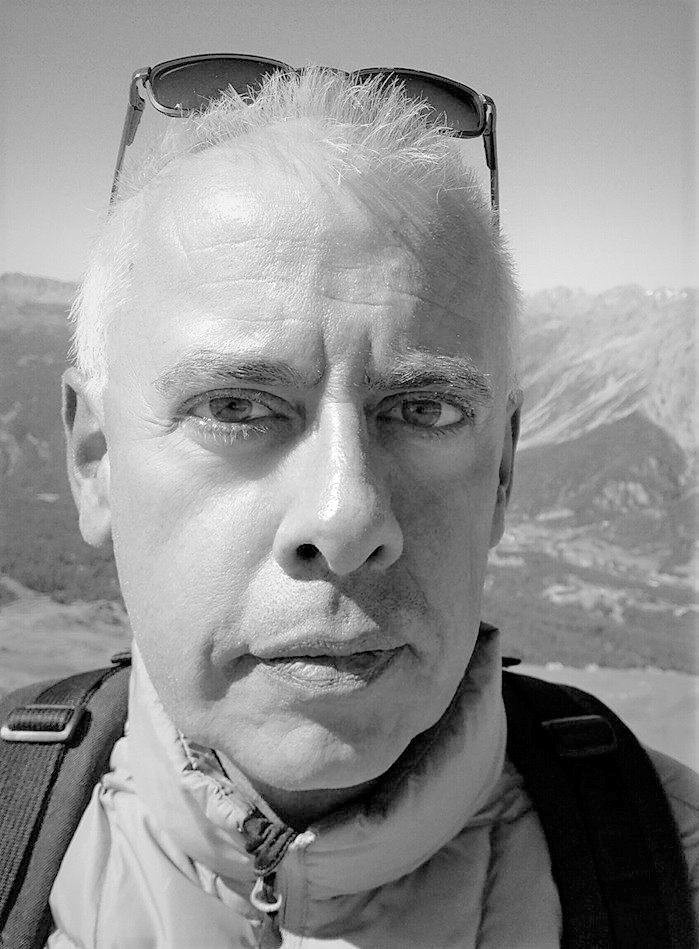}}]{Marcello Chiaberge} is currently Associate Professor within the Department of Electronics and Telecommunications, Politecnico di Torino, Turin, Italy. He is also the Co-Director of the Mechatronics Lab, Politecnico di Torino
(\url{www.lim.polito.it}), Turin, and the Director and the Principal Investigator of the new Centre for Service Robotics (PIC4SeR, \url{https://pic4ser.polito.it/}), Turin. He has authored more than 100 articles accepted in international conferences and journals,
and he is the coauthor of nine international patents. His research interests include
hardware implementation of neural networks and fuzzy systems and the design and implementation of reconfigurable real-time computing architectures. \end{IEEEbiography}
\vfill

% that's all folks
\end{document}